\documentclass[prd,preprintnumbers,amsmath,amssymb,floatfix]{revtex4}
\usepackage{amsmath,amsfonts,latexsym,amssymb,graphicx,graphics,epsfig,subfigure,color,makeidx,cases}
\usepackage{xcolor,diagbox,enumitem}
\usepackage{multirow}
\usepackage[colorlinks,linkcolor=blue,anchorcolor=black,citecolor=green,urlcolor=black]{hyperref}
\usepackage{mathrsfs}

\begin{document}
\title{Pseudospectrum of Braneworld Perturbations}

\author{Hai-Long Jia$^{a}$$^{b}$\footnote{jiahl2024@lzu.edu.cn}}
\author{Wen-Di Guo$^{a}$$^{b}$\footnote{guowd@lzu.edu.cn}}
\author{Yun-Tao Gu$^{a}$$^{b}$\footnote{guyt2024@lzu.edu.cn}}
\author{Yu-Xiao Liu$^{a}$$^{b}$\footnote{liuyx@lzu.edu.cn, corresponding author}}

\affiliation{
$^{a}$Lanzhou Center for Theoretical Physics, 
    Key Laboratory of Theoretical Physics of Gansu Province, 
    Key Laboratory of Quantum Theory and Applications of MoE, 
    Gansu Provincial Research Center for Basic Disciplines of Quantum Physics,
    Lanzhou University, Lanzhou 730000, China \\
$^{b}$Institute of Theoretical Physics \& Research Center of Gravitation,
    School of Physical Science and Technology, 
    Lanzhou University, Lanzhou 730000, China \\
}

\begin{abstract}
    
    Pseudospectral analysis provides a powerful way to probe the spectral
    stability of non-self-adjoint operators and has been widely used in black hole
    physics, but its application to braneworld scenarios has not yet been explored.
    In this work, we apply this method to tensor gravitational perturbations in a
    representative scalar-field-generated thick brane background. To the best
    of our knowledge, we provide the first hyperboloidal formulation of
    braneworld perturbations and propose a height-function gauge adapted to the
    warped geometry. This construction converts the outgoing boundary conditions
    of quasinormal modes into regularity conditions at finite compactified
    boundaries and recasts the perturbation equation as a first-order system
    generated by a non-self-adjoint hyperboloidal evolution operator. With the
    corresponding energy norm, we compute the condition numbers and
    pseudospectra of the localized graviton zero mode and the quasinormal-mode
    spectrum. We find that the condition numbers grow rapidly along the overtone
    sequence and that the corresponding pseudospectral contours develop broad,
    connected structures in the high-overtone region. These results show a
    strongly mode-dependent spectral sensitivity: among the damped modes analyzed,
    the higher overtones are less robust than the fundamental mode. The zero mode
    also has a larger condition number than the fundamental mode, indicating
    stronger local first-order sensitivity.
    These diagnostics characterize sensitivity to generic norm-bounded operator
    perturbations. Relating that sensitivity to a specific braneworld deformation
    requires the corresponding self-consistent perturbation constraints.

\end{abstract}

\maketitle
\tableofcontents 
\section{Introduction}\label{sec:introduction}

The braneworld paradigm describes our observable universe as a
four-dimensional hypersurface embedded in a higher-dimensional spacetime, with
ordinary matter confined to the brane while gravity can probe the bulk. Its
modern development is rooted in Dirichlet branes (D-branes)~\cite{Polchinski:1995mt}, large extra
dimensions~\cite{Arkani-Hamed:1998jmv,Antoniadis:1998ig}, and warped
geometries~\cite{Randall:1999ee,Randall:1999vf}. For a braneworld model to be
phenomenologically viable, it must reproduce four-dimensional gravity on the
scales probed by current experiments. In standard warped
constructions, gravity is localized by a normalizable graviton zero mode,
whereas quasi-localized scenarios such as the
Gregory-Rubakov-Sibiryakov (GRS) and Dvali-Gabadadze-Porrati (DGP)
models~\cite{Gregory:2000jc,Dvali:2000rv} produce an effective four-dimensional
regime through a long-lived graviton resonance. Beyond this low-energy limit,
the massive gravitational sector determines how perturbations propagate,
decay, and leak into the bulk.

In scalar-field-generated thick branes~\cite{DeWolfe:1999cp,Gremm:1999pj,
Csaki:2000fc,Gremm:2000dj}, the transverse-traceless tensor sector reduces,
after separation of the brane coordinates, to a master wave equation in time
and the extra dimension. The localized graviton zero mode ensures the
recovery of four-dimensional gravity, while the massive sector contains
Kaluza-Klein (KK) excitations~\cite{Kaluza:1921tu,
klein1926quantum,Overduin:1997sri}. Reviews of thick
brane gravity and localization can be found in
Refs.~\cite{Dzhunushaliev:2009va,Maartens:2010ar,Liu:2017gcn}. Depending on
the spectral formulation and boundary conditions, the massive sector may be
described by continuum modes, resonant states, or quasinormal modes (QNMs).
Early resonance studies in thin-brane and quasi-localized gravity
models~\cite{Csaki:2000pp,Brevik:2002yj,Clarkson:2005mg} were followed by analyses of
matter and gravitational resonances in thick
branes~\cite{Liu:2009ve,Liu:2009uca,Liu:2011wi}. Related de Sitter and two-field
brane resonances were investigated in Refs.~\cite{Araujo:2011fm,Cruz:2013uwa},
with gravity and fermion resonance spectra also studied on Bloch
branes~\cite{Xie:2013Bloch}. Gravitational resonances have further been explored
in branes generated by noncanonical scalar fields~\cite{Zhong:2014Noncanonical}
and in $f(R)$ and Eddington-inspired Born-Infeld
branes~\cite{Xu:2015Structure,Fu:2014EiBI}. Related developments include
warped $f(R)$ perturbations, fermion resonances, and gravitational
echoes~\cite{Zhong:2016iko,Zhu:2023tzx,Zhu:2024gvl}. 
The QNM formulation provides a complementary way to characterize dissipative
gravitational excitations. It was developed in Randall-Sundrum-type
setups~\cite{Seahra:2005wk,Seahra:2005iq,Chung:2015mna} and has recently been applied
to thick brane ringing and
resonance-QNM comparisons~\cite{Tan:2022vfe,Tan:2023cra}. Quasibound and QNM spectra in modified or
finite-domain thick branes were considered in Refs.~\cite{Tan:2024url,
Jia:2024pdk}, while graviscalar, double-brane, and de Sitter extensions were
studied in Refs.~\cite{Tan:2024aym,Tan:2024qij,Jia:2024sdk}. Recent analyses
have also addressed scalar-gravitational and $f(R)$ thick brane
QNMs~\cite{Deng:2025hfn,E:2025kic}. Such characteristic modes are connected with
possible corrections to the four-dimensional Newtonian
potential~\cite{Araujo:2011fm,Callin:2004py,Guo:2010az} and with gravitational-wave
phenomenology~\cite{Konoplya:2023fmh,Caprini:2018mtu}. The stability of this
characteristic spectrum is therefore a central ingredient in assessing the
dynamical robustness of braneworld scenarios.

Previous studies of braneworld spectra have mainly focused on the calculation
of characteristic modes and on their possible physical imprints.
More recently, the stability of these modes under background or effective
potential perturbations has begun to attract attention. In the
flat thick brane model, perturbations can lead to overtone or
fundamental-mode instabilities,
although their direct imprint in the time-domain waveform may be strongly
suppressed~\cite{Jia:2026vnx}. In the de Sitter thick brane model studied in
Ref.~\cite{Jia:2024sdk}, where the unperturbed QNM frequencies are purely imaginary,
perturbations can generate nonzero real parts and hence produce transient
oscillatory behavior. These results show that the characteristic spectrum of a
braneworld background is not merely a fixed set of frequencies, but an object
whose stability requires a systematic analysis.

Direct calculations with prescribed deformations of the brane background or
the effective potential determine how those specific deformations move the QNM
spectrum. Pseudospectral analysis complements this approach by quantifying the
largest spectral response allowed by an operator-norm bound, without selecting
a deformation family in advance. This distinction is particularly relevant to
QNMs: once outgoing-wave boundary conditions are imposed, the spectral problem
is governed by a non-self-adjoint operator, for which small operator
perturbations can produce large eigenvalue displacements.

The appropriate tools for this purpose are the condition number and the
pseudospectrum~\cite{Trefethen:2005}. The condition number gives a local,
first-order measure of the sensitivity of an individual eigenvalue, while the
pseudospectrum is the union of spectra generated by all operator perturbations
within a prescribed norm bound. Pseudospectral methods have already revealed
important aspects of QNM instability in black hole
physics~\cite{Jaramillo:2020tuu,Cheung:2021bol,Jaramillo:2021tmt,Berti:2022xfj}.
To the best of our knowledge, such an analysis has not yet been
applied to braneworld perturbations. This motivates the central question of the
present work: how stable are the localized zero mode and the QNM spectrum of a
thick brane in the pseudospectral sense?

In this paper, we address this question for tensor perturbations of a
representative scalar-field-generated thick brane. We first construct a
hyperboloidal compactification of the perturbation equation. This formulation maps the
noncompact extra dimension to a finite interval and converts the outgoing QNM
boundary conditions into regularity conditions at the compactified endpoints.
The problem is then written as a first-order evolution system generated by a
non-self-adjoint operator. Using the energy norm associated with this
hyperboloidal system, we define the condition numbers and pseudospectra of the
characteristic modes and compute them with a Chebyshev spectral method.

Our analysis shows that the pseudospectral sensitivity is strongly mode
dependent. The QNM overtones acquire rapidly increasing condition numbers, and
the contours in the high-overtone region develop broad and connected
structures. Among the damped modes displayed, the higher overtones are
therefore spectrally less robust than the fundamental QNM. The localized zero
mode also has a larger condition number than the fundamental QNM, indicating
stronger local first-order sensitivity. These comparisons quantify generic
operator-level sensitivity in the chosen norm. Mode trajectories produced by a
self-consistent braneworld deformation constitute a more restricted problem.

The paper is organized as follows. In
Sec.~\ref{sec:hyperboloidal-compactification}, we review the tensor perturbation
equation of the thick brane and construct the hyperboloidal compactification
used in the spectral analysis. In Sec.~\ref{sec:spectral-sensitivity}, we
introduce the condition number, the energy-norm pseudospectrum, and the
Chebyshev spectral method. We then present the numerical results for the
characteristic spectrum, condition numbers, and pseudospectral structures.
Finally, Sec.~\ref{sec:conclusion} summarizes the main conclusions and discusses
possible extensions of this work.

\section{Hyperboloidal Compactification for Braneworld Perturbations}\label{sec:hyperboloidal-compactification}

\subsection{Braneworld Background and Gravitational Perturbations}\label{sec:brane-background}

To prepare for the hyperboloidal treatment and the subsequent pseudospectral
analysis, we first identify the perturbation sector that leads to a well-defined
wave operator on an unbounded extra-dimensional domain. Among the linear
perturbation channels of a thick brane, the transverse-traceless tensor sector
is the most suitable starting point. It directly probes the gravitational
localization properties of the warped geometry and, at linear order, decouples
from scalar-field perturbations. The dynamics can therefore be reduced to a
single master equation with an effective potential.
This makes it an ideal setting for studying resonances, quasinormal frequencies,
and, more generally, the spectral sensitivity of the associated non-self-adjoint
wave operator.

We consider a five-dimensional thick brane generated by a real bulk scalar field
$\phi$, described by the action~\cite{DeWolfe:1999cp,Gremm:1999pj}
\begin{equation}
S=\int d^4x\,dy\,\sqrt{-g}\left[
\frac{1}{2\kappa^2}R
-\frac{1}{2}\nabla_M\phi\nabla^M\phi
-V(\phi)
\right].
\end{equation}
Here $\kappa$ denotes the five-dimensional gravitational constant and $V(\phi)$
is the self-interaction potential of the scalar field. In this work, we set
$\kappa=1$. We restrict attention to static flat brane configurations preserving
four-dimensional Poincar\'e invariance, for which the background metric takes the form
\begin{equation}
ds^2=e^{2A(y)}\eta_{\mu\nu}dx^\mu dx^\nu+dy^2,
\end{equation}
where $A(y)$ is the warp factor, $\eta_{\mu\nu}=\mathrm{diag}(-1,1,1,1)$ is the
four-dimensional Minkowski metric, and $y$ denotes the physical coordinate of
the extra dimension. The scalar field and the warp factor depend only on $y$.

The background equations obtained from the Einstein-scalar system are
\begin{align}
    \label{phi-field-equation}
    \text{matter}&:& \phi''+4 A' \phi' &= \frac{\partial V(\phi)}{\partial \phi}, \\
    \label{thick-field-equation-munu}
    (\mu,\nu)&:&  6A'^2+ 3A''  &= -\frac{1}{2}\phi'^2- V(\phi) ,\\
    \label{thick-field-equation-55}
    (5,5)&:&  6A'^2 &= \frac{1}{2}\phi'^2- V(\phi), 
\end{align}
where a prime denotes differentiation with respect to $y$.
These equations determine the background thick brane solution, often through
the first-order formalism~\cite{DeWolfe:1999cp,Afonso:2006gi}.

For the perturbation analysis, it is convenient to introduce 
the conformal coordinate $z$ defined by
\begin{equation}
dz=e^{-A(y)}dy.
\end{equation}
In terms of $z$, the background metric takes a conformally flat form, 
which simplifies the tensor perturbation equations.
We consider the transverse-traceless perturbation of the metric, 
\begin{equation}
ds^2=e^{2A(z)}\left[(\eta_{\mu\nu}+h_{\mu\nu})dx^\mu dx^\nu+dz^2\right],
\end{equation}
subject to
\begin{equation}
\partial^\mu h_{\mu\nu}=0,
\qquad
\eta^{\mu\nu}h_{\mu\nu}=0.
\end{equation}
Because of these conditions, the tensor sector is gauge invariant at linear
order and evolves independently of the scalar perturbation. The linearized
Einstein equation then reduces to
\begin{equation}
\left[
\eta^{\alpha\beta}\partial_\alpha\partial_\beta
+\partial_z^2
+3 (\partial_z A) \partial_z
\right]h_{\mu\nu}=0.
\end{equation}
To separate the four-dimensional and extra-dimensional dynamics, we write
\begin{equation}
h_{\mu\nu}(x^\lambda,z) = \varepsilon_{\mu\nu}\, e^{-ip_j x^j}\, e^{-\frac{3}{2}A(z)}\, \Psi(t,z),
\end{equation}
where $\varepsilon_{\mu\nu}$ is a constant polarization tensor and
$p^2=\delta^{ij}p_i p_j$ is the squared magnitude of the spatial momentum along the brane. Substituting
this ansatz into the perturbation equation yields the master wave equation
\begin{equation}
\left[ -\partial_t^2 +\partial_z^2-V(z)-p^2\right]\Psi(t,z)=0,
\end{equation}
with the effective potential
\begin{equation}
    \label{effective-potential-z}
    V(z) =\frac{3}{2}\partial_{z}^{2}A + \frac{9}{4}\left(\partial_{z}A\right)^2.
\end{equation}

Thus, for each brane momentum and polarization tensor, the perturbation sector
is reduced to a $(1+1)$-dimensional master equation in $(t,z)$ coordinates, 
with all background information encoded in the effective potential $V(z)$.

Assuming harmonic time dependence,
\begin{equation}
\Psi(t,z)=e^{-i\omega t}\psi(z),
\end{equation}
the master equation becomes
\begin{equation}
\left[-\partial_z^2+V(z)\right]\psi(z)=m^2\psi(z), \qquad m^2=\omega^2-p^2.
\end{equation}
This Schr\"odinger-type equation provides the basic spectral problem associated
with tensor perturbations. Its Hamiltonian admits the factorization
\begin{equation}
-\partial_z^2+V(z) = \left(\partial_z+\frac{3}{2}\partial_{z}A \right)
\left(-\partial_z+\frac{3}{2}\partial_{z}A \right),
\end{equation}
which shows that the spectrum is non-negative in the standard self-adjoint
setting and, in particular, excludes tachyonic tensor instabilities. 
The corresponding graviton zero mode is
\begin{equation}
\psi_0(z)\propto e^{\frac{3}{2}A(z)},
\end{equation}
and its normalizability determines whether four-dimensional gravity can be
localized on the brane. Beyond the zero mode, the system generally possesses a
continuum of massive KK modes. For typical thick brane backgrounds, the effective
potential has a volcano-like profile: it is localized near the brane and
decays in the asymptotic region~\cite{DeWolfe:1999cp,Gremm:1999pj,
Csaki:2000fc}. Such a potential does not usually support many
normalizable bound states, but it may give rise to resonant modes~\cite{Xie:2013Bloch,Zhong:2014Noncanonical,Xu:2015Structure,Fu:2014EiBI}, 
namely, quasi-localized massive excitations that remain trapped near the brane for a
long time before leaking into the bulk. This mechanism appears in
quasi-localized gravity~\cite{Csaki:2000pp}, in thick brane
resonance spectra~\cite{Liu:2009ve,Araujo:2011fm,Cruz:2013uwa}, and in recent comparisons
between resonances and QNMs~\cite{Tan:2023cra}. These resonances are of
particular
interest because they represent the spectral imprint of the warped geometry
beyond the localized graviton zero mode~\cite{Seahra:2005wk,Tan:2023cra}.

At this stage it is important to distinguish two related but conceptually
different spectral problems. 
In the standard self-adjoint formulation, the spectrum consists of a
normalizable zero mode and a continuum of massive KK scattering states~\cite{DeWolfe:1999cp,Gremm:1999pj,Csaki:2000fc}. 
Instead, if one imposes purely outgoing conditions in the asymptotic region, 
the problem becomes non-self-adjoint and its discrete complex frequencies 
correspond to QNMs~\cite{Seahra:2005wk,Seahra:2005iq,
Tan:2022vfe,Tan:2023cra}. 
Since the pseudospectrum is designed to probe the stability and sensitivity of
non-self-adjoint operators, the outgoing-wave formulation is the one relevant for
the present work~\cite{Trefethen:2005,Jaramillo:2020tuu}. 
The hyperboloidal framework introduced in the
next subsection provides a natural way to implement these radiative boundary conditions 
and to reformulate the perturbation equation on a compactified domain.

As a representative example, we will later specialize to the thick brane
solution~\cite{DeWolfe:1999cp,Gremm:1999pj}
\begin{align}
\label{solution}
A(y) &= -c \ln\!\big[\cosh(ky)\big], \\
\phi(y) &= \sqrt{3c}\,\arcsin\!\big[\tanh(ky)\big], \\
V(\phi) &= \frac{3k^2 c}{4}
\left[
1-4c+(1+4c)\cos\!\left(\frac{2\phi}{\sqrt{3c}}\right)
\right],
\end{align}
where $k$ sets the characteristic energy scale of the five-dimensional bulk 
and controls the brane thickness, 
while $c>0$ is a dimensionless parameter labeling the family of solutions.
This background is regular, asymptotically warped, and supports a normalizable
graviton zero mode~\cite{DeWolfe:1999cp,Gremm:1999pj}.
Moreover, its tensor potential exhibits the typical
localized barrier structure needed for studying resonances and quasinormal
spectra, making it a convenient benchmark model for the pseudospectral analysis.

\subsection{Geometric Foundations of the Hyperboloidal Approach}\label{sec:geometric-foundations}

For the geometric discussion, it is convenient to suppress the brane momentum by
setting $p=0$. Since $p^2$ enters as a zeroth-order algebraic term rather than a
derivative term, it does not affect the principal part of the master equation
and therefore does not modify its asymptotic causal structure. The wave equation obtained in the
previous subsection then reduces to

\begin{equation}
    \label{wave-equation}
\left[-\partial_t^2+\partial_z^2-V(z)\right]\Psi(t,z)=0,
\end{equation}
posed on the noncompact domain $z\in(-\infty,\infty)$. For the class of thick
brane backgrounds considered here, the effective potential $V(z)$ is localized
near the brane core and decays in the asymptotic region. Consequently, the
large-$|z|$ propagation of the field is controlled by the principal part of the
operator, namely the $(1+1)$-dimensional wave operator associated with the
auxiliary metric
\begin{equation}
    \label{auxiliary-metric}
ds_2^2=-dt^2+dz^2.
\end{equation}
From this perspective, the hyperboloidal approach may be understood as a geometric
reformulation of the asymptotic wave problem~\cite{Jaramillo:2020tuu,Zenginoglu:2007jw}. Instead of evolving the field on
standard Cauchy slices $t=\mathrm{const}$, which terminate at spatial infinity,
one introduces spacelike slices that asymptote to future null infinity. In this
way, the radiative character of the physical solution is incorporated into the
foliation itself, rather than imposed externally through approximate boundary
conditions at a large but finite radius. 

\begin{figure*}[htb]
    \begin{center}
    \includegraphics[width=6.0cm]{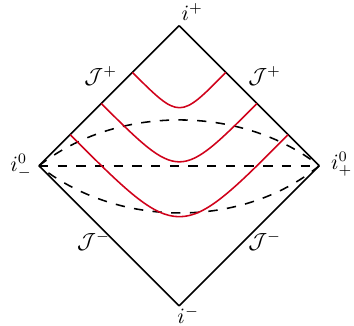}
    \end{center}
    \caption{Penrose diagram of the braneworld spacetime, 
    showing the standard time slices (dashed lines) and 
    the future hyperboloidal time slices (solid red lines).}\label{figure-1}
\end{figure*}

To implement this idea, we replace the original time coordinate $t$ by a new time function
\begin{equation}
    \label{time-transformation}
\tau=t - h(z),
\end{equation}
where $h(z)$ is the height function. In the coordinates $(\tau,z)$, the
auxiliary metric~\eqref{auxiliary-metric} becomes
\begin{equation}
ds_2^2=-d\tau^2-2H\,d\tau\,dz+\left(1-H^2\right)dz^2,
\end{equation}
where $H(z)=\frac{dh}{dz}$ is the boost function associated with the slicing. 
In order for the hypersurfaces $\tau=\mathrm{const}$ to 
define a regular hyperboloidal foliation, the boost
function must satisfy the following requirements.

\begin{itemize}
    \item First, the slices should remain spacelike for finite $z$, which implies
    \begin{equation}
        \label{boost-condition-1}
    |H(z)|<1.
    \end{equation}

    \item Second, the slices must approach the outgoing null directions asymptotically, namely,
    \begin{equation}
        \label{boost-condition-2}
    H(z)\to \pm 1 \quad (z\to \pm \infty). 
    \end{equation}
    Thus the hyperboloidal slices bend upward toward future null infinity at both
    ends of the physical domain, as illustrated in Fig.~\ref{figure-1}.

    \item Finally, one usually imposes
    \begin{equation}
        \label{boost-condition-3}
    \frac{dH}{dz}\to 0  \quad (z\to\pm\infty),
    \end{equation}
    so that the approach to the null directions has a vanishing asymptotic slope.
    Moreover, as will be seen later, this condition prevents the appearance of an additional divergent term,
    $i\omega \partial_z H$, in the Schr\"odinger-like equation.
\end{itemize}

The usefulness of the height function becomes especially clear at the level of
the frequency-domain problem. Assuming harmonic time dependence,
\begin{equation}
\Psi(t,z)=e^{-i\omega t}\psi(z),
\end{equation}
the master equation reduces to the Schr\"odinger-like equation
\begin{equation}
\left(\frac{d^2}{dz^2}+\omega^2-V(z)\right)\psi(z)=0.
\end{equation} 
For the representative thick brane solution~\eqref{solution},
the effective potential $V(z)$ vanishes at the asymptotic boundaries.
Therefore, the asymptotic solution takes the form
$\psi \to C_1 e^{i\omega z}+C_2 e^{-i\omega z}$.
Accordingly, the QNM boundary condition is the purely outgoing one,
\begin{equation}
\psi(z)\sim e^{\pm i\omega z} \quad (z\to \pm\infty).
\end{equation}
For damped modes with $\mathrm{Im}\,\omega<0$, these outgoing solutions grow
exponentially in magnitude on the standard Cauchy slices 
(see the black dotted line in Fig.~\ref{figure-1}). 
This is the familiar difficulty of the QNM problem on an unbounded domain: 
the physically relevant radiative boundary condition is represented by 
asymptotically divergent mode functions.

Within the hyperboloidal framework, however, the time transformation in
Eq.~\eqref{time-transformation} naturally induces a rescaling of the frequency-domain
field, $\tilde{\psi}=e^{-i\omega h}\psi$, so that
\begin{equation}
\Psi(\tau,z)=e^{-i\omega \tau}\tilde{\psi}(z).
\end{equation}
Substituting this expression into the wave equation yields
\begin{equation}
\left[ \frac{d^2}{dz^2} +2i\omega H(z)\frac{d}{dz} +\omega^2\bigl(1-H(z)^2\bigr)
+i\omega \frac{dH}{dz} -V(z) \right]\tilde{\psi}(z)=0.
\end{equation}
Using the asymptotic properties~\eqref{boost-condition-2} and~\eqref{boost-condition-3}
of the boost function, the leading equation at either asymptotic end reduces to
\begin{equation}
\left(\frac{d^2}{dz^2}\pm 2i\omega \frac{d}{dz}\right)\tilde{\psi}=0, \qquad z\to \pm \infty.
\end{equation}
Hence
\begin{equation}
\tilde{\psi}(z) \sim C_1 e^{\mp 2i\omega z}+C_2, \qquad z\to \pm \infty. 
\end{equation}
The outgoing branch therefore approaches a finite constant at each asymptotic
end. The hyperboloidal transformation thus removes the exponential growth of
the outgoing QNM eigenfunctions on standard Cauchy slices. In the compactified
formulation, outgoing boundary conditions are encoded through an appropriate
regularity requirement on the operator domain, rather than through boundedness
alone.

The introduction of hyperboloidal time slices already captures the correct
outgoing asymptotics, but the spatial domain is still noncompact. To bring the
asymptotic boundaries to finite coordinate locations, one further introduces a
compactifying coordinate $u$ through
\begin{equation}
    \label{spatial-transformation}
z=\frac{u}{\Omega(u)},
\end{equation}
where the compactification factor $\Omega(u)$ is positive in the interior
and vanishes at the endpoints of the compact domain. 
For definiteness, one may take $u\in[u_-,u_+]$ with
\begin{equation}
    \label{compactification-condition}
\Omega(u_{\pm})=0,  \qquad \Omega(u)>0 \quad \text{for} \; u_-< u <u_+.
\end{equation}
The two asymptotic regions $z\to\pm\infty$ are then mapped to the finite
locations $u=u_{\pm}$. 
Differentiating Eq.~\eqref{spatial-transformation} gives
\begin{equation}
dz=\frac{J(u)}{\Omega(u)^2}\,du,
\end{equation}
where $J(u)=\Omega(u)-u\,\partial_u\Omega$.
Accordingly, the auxiliary metric can be rewritten in the compactified coordinates $(\tau,u)$.

A key point of the hyperboloidal compactification is that the metric becomes
singular only as a coordinate artifact. 
This coordinate divergence is removed by the conformal rescaling $\check g_{ab}=\Omega^2 g_{ab}$,
under which the rescaled auxiliary line element is
\begin{equation}
d\check s_2^2 = -\Omega(u)^2 d\tau^2 -2H(u)J(u)\,d\tau\,du
+\left[1-H(u)^2\right]\frac{J(u)^2}{\Omega(u)^2}du^2.
\end{equation}

To ensure that the conformally rescaled metric has a finite, regular limit at
the compactified boundary, the asymptotic behavior of the foliation must be
appropriately related to the compactification factor.

The regularity requirement is
\begin{equation}
1-H(u)^2=\mathcal{O}\!\left(\Omega(u)^2\right), \qquad u\to u_{\pm}.
\end{equation}
Geometrically, this expresses the fact that the slices become asymptotically null
at the same rate at which the infinite domain is compactified. Under this condition,
the conformal boundaries $u=u_{\pm}$ represent future null infinity in fixed
coordinate positions, a construction often referred to as \textit{scri-fixing}~\cite{Zenginoglu:2007jw}.

It is important to emphasize that conformal compactification does not alter the
asymptotic null character of the hyperboloidal slices. 
Instead, its role is to remove the coordinate singularity associated with the infinite physical domain
and to render the compactified metric regular at the boundaries. In this sense,
hyperboloidal slicing and conformal compactification are complementary: the
former incorporates the correct outgoing asymptotics into the time foliation,
whereas the latter maps the asymptotic endpoints onto a finite coordinate
interval. This geometric construction has a direct analytic consequence. In the
original coordinates, outgoing boundary conditions must be imposed as
$z\to\pm\infty$ by hand, typically after truncating the domain at a large but
finite radius. In the hyperboloidal compactified picture, by contrast, the
boundaries of the finite interval correspond to characteristic outflow
boundaries of the evolution problem. As a result, the radiative behavior of the
solution is encoded geometrically rather than enforced through artificial outer
boundary conditions, which is precisely what makes the hyperboloidal approach so
well suited to the analysis of QNMs and pseudospectra.

In summary, the hyperboloidal framework provides a natural geometric setting for
the braneworld perturbation problem. The height function $h(z)$ changes the time
foliation in such a way that the purely outgoing boundary conditions of the QNM
problem are replaced by a regularity condition at the asymptotic ends, while the
compactification $z=u/\Omega(u)$, together with the conformal factor
$\Omega(u)$, maps the unbounded physical domain to a finite interval with
regular boundaries. Although the zeroth-order terms $V(z)$ and $p^2$ determine
the detailed scattering and resonance structure of the wave operator, they do
not modify its asymptotic causal behavior. As a result, the geometric
construction developed above makes it possible to regularize the perturbation
equation on a compact domain without sacrificing its physically relevant
outgoing asymptotics. In the following subsections, we specify the gauge functions and derive the
compactified hyperboloidal form of the master equation for $\Psi(t,z)$. 

\subsection{Gauge Choice for the Height and Compactification Functions}\label{sec:gauge-choice}

At this stage, the hyperboloidal construction still contains a considerable
amount of gauge freedom. Indeed, the geometric requirements discussed in the
previous subsection do not uniquely determine either the height function
$h(z)$ or the compactification factor $\Omega(u)$. In practical calculations,
it is therefore convenient to fix this freedom by adopting a gauge that is
compatible with the symmetry of the braneworld background and at the same time
keeps the coefficients of the compactified wave equation as simple as possible.

We begin with the choice of the height function. A convenient general strategy
is to construct the boost function $H(z)$ first and then recover the height
function from the relation
\begin{equation}
h(z)=\int^z H(\zeta)\,d\zeta,
\end{equation}
with the integration constant fixed. The boost function must
satisfy the three conditions in
Eqs.~\eqref{boost-condition-1}--\eqref{boost-condition-3}
stated in the previous subsection.
To encode these requirements in a convenient way, 
we introduce a positive ``decay function'' $f(z)$ through
\begin{equation}
1-H(z)^2=f(z).
\end{equation}

For a reflection-symmetric thick brane background, we choose $f(z)$ to be even.
The relation above fixes only $H^2$, so the parity of $H$ must be specified
separately. To obtain a reflection-symmetric, future-directed foliation with
$H\to-1$ at the left endpoint and $H\to+1$ at the right endpoint, we select the
odd branch

\begin{equation}
H(z)=\mathrm{sgn}(z)\sqrt{1-f(z)}.
\end{equation}
We require $f(z)$ to satisfy
\begin{equation}
0<f(z)\le 1, \qquad f(z)\to 0, \qquad \frac{df}{dz}\to 0 \qquad (z\to\pm\infty).
\end{equation}

These conditions guarantee the spacelike inequality and the required
asymptotic limits of the selected branch. Regularity where $f=1$ must also be
checked for a particular choice; the explicit function used below gives a
regular $H$ at $z=0$. In this parametrization, the decay of $f(z)$ directly
controls how rapidly the hyperboloidal slices approach null infinity.

A particularly natural choice in braneworld models is to relate the decay
function to the warp exponent,
\begin{equation}
f(z)=e^{2\eta A(z)},  
\qquad
\eta>0,
\end{equation}
where $\eta$ is a gauge parameter. 
The reason is that the warped geometry controls both the localization of the
brane and the asymptotic behavior of the effective potential. For the class of
symmetric backgrounds considered here, we normalize the warp exponent by
$A(0)=0$ and have $A(z)\leq0$, with $A(z)\to-\infty$ as
$|z|\to\infty$. Consequently,
$0<e^{2\eta A(z)}\leq1$, and the function decays to zero at both ends. The
representative solution below also satisfies
$d(e^{2\eta A})/dz\to0$, so all conditions imposed on $f$ are
explicitly satisfied. In what follows, we adopt the simplest choice $\eta=1$.
More generally, different values of $\eta$ correspond to different but equally
admissible gauges,
changing the detailed shape of the slices without affecting the underlying
physics.

For the analytically convenient thick brane
solution~\cite{DeWolfe:1999cp,Gremm:1999pj} with $c=1$, 
\begin{equation}
e^{2A(z)}=\frac{1}{1+k^2z^2},
\end{equation}
one obtains
\begin{equation}
H(z)=\frac{kz}{\sqrt{1+k^2z^2}}.
\end{equation}
Integrating this expression yields
\begin{equation}
    \label{height-ansatz}
h(z)=\frac{1}{k}\left(\sqrt{1+k^2z^2}-1\right),
\end{equation}
where the additive constant has been chosen so that $h(0)=0$. This choice makes
the hyperboloidal slicing completely explicit and leads to particularly simple
expressions in the subsequent analysis. For more general thick brane
backgrounds, the same construction remains valid, although the integral for
$h(z)$ may no longer be elementary.

We next turn to the compactification. 
Once the hyperboloidal slicing has been fixed, 
the remaining freedom lies in the map that brings the infinite
$z$-domain to a finite interval. 
According to condition~\eqref{compactification-condition}, 
the compactification factor $\Omega(u)$ must vanish at the boundaries of the
compact interval and remain positive in the interior. For symmetric backgrounds,
it is again natural to choose an even function of $u$. 
In addition, boundary regularity requires that 
$J(u)$ remain finite and nonzero at the compactified boundaries. 
This guarantees that the coordinate 
singularity associated with $z\to\pm\infty$ is completely absorbed by the
conformal factor, while the remaining coefficients of the compactified equation
stay regular.

Taking into account the reflection symmetry of the warp factor, we adopt the
simplest even quadratic ansatz
\begin{equation}
    \label{omega-ansatz}
\Omega(u)=\frac{1}{2}(1-k^2u^2) =\frac{1}{2}(1+ku)(1-ku).
\end{equation}
The compactified boundaries are therefore located at $u_\pm=\pm \frac{1}{k}$. 
With this choice, we find
\begin{equation}
J(u)=\Omega(u)-u \partial_u\Omega =\frac{1}{2}(1+k^2u^2),
\end{equation}
and hence $J\! \left(\pm\frac{1}{k}\right)=1$. 
This choice is useful because the Jacobian factor remains finite and simple at
the boundaries. 
Moreover, substituting $z=u/\Omega(u)$ into the boost function gives
\begin{equation}
1-H(u)^2 =\frac{(1-k^2u^2)^2}{(1+k^2u^2)^2}.
\end{equation}
Combined with Eq.~\eqref{omega-ansatz}, 
we immediately obtain
\begin{equation}
1-H(u)^2=\mathcal{O}\!\left(\Omega(u)^2\right) 
    \qquad \text{as} \;\; u\to u_\pm,
\end{equation}
so that the regularity condition required for conformal compactification is
indeed satisfied. In this sense, the above choice of $\Omega(u)$ is fully
compatible with the previously chosen hyperboloidal slicing.

The above choice is not unique. Other decay functions $f(z)$ and
compactification factors $\Omega(u)$ may be adopted, provided that the same
asymptotic and regularity conditions are respected. We choose the present gauge
because it respects reflection symmetry and keeps the transformed coefficients
simple. In particular, it gives closed-form expressions for $H(z)$, $h(z)$,
and $J(u)$; substituting these expressions into the wave equation yields regular
rational coefficient functions on the compact interval.

These choices do not change the background or the QNM
boundary-value problem; their role is to provide a regular,
computationally convenient representation. The pseudospectrum remains
norm-dependent, as discussed below; throughout this work it is evaluated in
this fixed gauge with the corresponding energy norm.

\subsection{Hyperboloidal Compactification of the Wave Equation}\label{sec:hyperboloidal-wave}

We now apply the geometric construction introduced above directly to the
braneworld perturbation equation. As in the previous subsection, we restrict
attention to the sector with $p=0$. 
Under the compactified hyperboloidal transformation
\begin{align}
    t &= \tau + h(u), \\
    z &= g(u),
\end{align}
where $h(u)\equiv h(g(u))$ and $g(u)$ is the hyperboloidal compactification map, 
the derivatives transform according to
\begin{equation}
    \partial_t=\partial_\tau,
    \qquad
    \partial_z=\frac{1}{g_u}\left(\partial_u-h_u\partial_\tau\right),
\end{equation}
where a subscript $u$ denotes differentiation with respect to $u$.
Therefore, the wave equation~\eqref{wave-equation} takes the form
\begin{equation}
    \label{wave-equation-re}
\left[\left(1-\left(\frac{h_u}{g_u}\right)^2\right)\partial_\tau^2
+ \frac{2}{g_u}\left(\frac{h_u}{g_u}\right)\partial_\tau\partial_u
+ \frac{1}{g_u} \partial_u\!\left(\frac{h_u}{g_u}\right) \partial_\tau \\
-\frac{1}{g_u}\partial_u\!\left(\frac{1}{g_u}\partial_u\right) + V(u)\right]\Psi(\tau,u)=0, 
\end{equation}
where $V(u)\equiv V(g(u))$ is the effective potential expressed in the
compactified coordinate by composition with $g$.
It is convenient to rewrite Eq.~\eqref{wave-equation-re} in a form adapted to
the compactified geometry. Multiplying by $|g_u|$, we obtain
\begin{equation}
    \label{wave-equation-regular}
    \left[
    w(u)\,\partial_\tau^2
    -\left(2\gamma(u)\partial_u+\partial_u\gamma(u)\right)\partial_\tau
    -\partial_u\!\left(\xi(u)\partial_u\right)
    +\hat{V}(u)
    \right]\Psi(\tau,u)=0,
\end{equation}
where
\begin{equation}
    \label{wpgamma-def}
    w(u)=\frac{g_u^2-h_u^2}{|g_u|},
    \qquad
    \xi(u)=\frac{1}{|g_u|},
    \qquad
    \gamma(u)=-\frac{h_u}{|g_u|},
    \qquad
    \hat{V}(u)=|g_u|\,V(u).
\end{equation}
Here the absolute value is introduced only to keep the formulas independent of
the orientation of the compactification map; for the gauge adopted below one has
$g_u>0$, so $|g_u|=g_u$. Equation~\eqref{wave-equation-regular} makes the
hyperboloidal structure more transparent: the second-order part in $u$ is
weighted by $\xi(u)$. Within the spatial operator
$2\gamma\partial_u+\partial_u\gamma$, the term
$2\gamma\partial_u$ is the transport part, while $\partial_u\gamma$ is the
accompanying multiplicative term. Together they arise from the hyperboloidal
shift and provide a direct source of non-self-adjointness in the outgoing
spectral formulation.

To cast the problem into an evolution form, we perform a first-order reduction in
time by introducing
\begin{equation}
    \mathcal{V}=
    \begin{pmatrix}
        \Psi \\
        \Pi
    \end{pmatrix},
    \qquad
    \Pi=\partial_\tau \Psi.
\end{equation}
Equation~\eqref{wave-equation-regular} is then equivalent to
\begin{equation}
    \label{first-order-equation}
    \partial_\tau \mathcal{V}=-iL\,\mathcal{V},
\end{equation}
with
\begin{equation}
    \label{L-operator}
    L=i
    \begin{pmatrix}
        0 & 1 \\
        L_1 & L_2
    \end{pmatrix}
    =
    i
    \begin{pmatrix}
        0 & 1 \\
        \dfrac{1}{w(u)}
        \left[\partial_u\!\left(\xi(u)\partial_u\right)-\hat{V}(u)\right]
        &
        \dfrac{1}{w(u)}
        \left[2\gamma(u)\partial_u+\partial_u\gamma(u)\right]
    \end{pmatrix}.
\end{equation}
In this form the perturbation problem is represented as a first-order evolution
system on a compact interval. Since the transformation $t=\tau+h(u)$ preserves
the stationary Killing flow, one still has $\partial_\tau=\partial_t$ at the
level of the generator. Consequently, the associated frequency parameter is the
same in both coordinate systems, and the hyperboloidal transformation does not
change the QNM spectrum itself.

Indeed, after Fourier decomposition with respect to $\tau$,
$\mathcal{V}(\tau,u)=e^{-i\omega\tau}\mathcal{V}(u)$, 
Eq.~\eqref{first-order-equation} reduces to the eigenvalue problem
\begin{equation}
    \label{eigenvalue-problem}
    L\,\mathcal{V}_n=\omega_n\,\mathcal{V}_n.
\end{equation}
Thus the QNM problem is reformulated as a spectral problem for the
non-self-adjoint operator $L$ on the compactified hyperboloidal domain.

The boundary conditions are also simplified in this framework. As discussed in
the previous subsection, the height function has already converted the outgoing
conditions at $z\to\pm\infty$ into regularity conditions at the compactified
endpoints. Moreover, for admissible compactifications one has $\xi(u_\pm)=0$ at
the endpoints $u_\pm$, so that they become singular Sturm-Liouville endpoints
of the spatial operator. No additional independent boundary conditions need be
imposed there: it is sufficient to require that the solution remain regular up
to the boundaries. The resulting eigenvalues $\omega_n$ are precisely the QNM
frequencies, and the corresponding regular eigenfunctions are the associated
mode functions in the hyperboloidal representation.

The geometric construction in the previous subsection has already fixed
$h(u)$ and $g(u)$. We now verify that this choice also gives regular coefficients
for the first-order spectral problem: $w(u)$ must remain positive in the
interior, $\gamma(u)$ must be regular on the closed interval, and $\xi(u)$ must
vanish regularly at the endpoints. For the representative thick brane solution, the
chosen gauge gives

\begin{align}
    h(u)&=\frac{2ku^2}{1-k^2u^2}, \\
    g(u)&=\frac{2u}{1-k^2u^2}.
\end{align}
The compactified boundaries are then located at $u_\pm=\pm \frac{1}{k}$,
and the corresponding coefficient functions become
\begin{equation}
    w(u)=\frac{2}{1+k^2u^2},
    \qquad
    \xi(u)=\frac{(1-k^2u^2)^2}{2(1+k^2u^2)},
    \qquad
    \gamma(u)=-\frac{2ku}{1+k^2u^2}.
\end{equation}
Hence $w(u)>0$ throughout the interior, $\gamma(u)$ is regular on the whole
compact interval, and $\xi(u_\pm)=0$. This explicitly verifies the coefficient
conditions required by the compactified spectral equation.
Furthermore, because the effective potential decays sufficiently fast in the
asymptotic region, $\hat{V}(u)$ is also regular at the compactified endpoints.
This makes the above gauge a particularly simple and effective choice for the
subsequent numerical and pseudospectral analysis.

\section{Spectral Sensitivity and Pseudospectrum of Braneworld Perturbations}\label{sec:spectral-sensitivity}

For generalized spectral problems, and in particular for non-self-adjoint
operators, the stability of the spectrum under perturbations is a central issue.
Two standard tools for quantifying such spectral stability are the condition
number of individual eigenvalues and the pseudospectrum of the operator. Both
concepts depend on the chosen norm, because an operator perturbation can have
different sizes in different norms. In the definitions below,
$\dagger$ denotes the adjoint, and $\|\cdot\|$ denotes the vector norm or its
induced operator norm, as appropriate, associated with an arbitrary but fixed inner product. We then specialize to the energy
norm~\cite{Jaramillo:2020tuu} associated with the hyperboloidal wave equation
before performing the braneworld calculation.

\subsection{Definitions of the Condition Number and Pseudospectrum}\label{sec:definitions}

Before turning to the numerical computation of pseudospectra, we first briefly
review the definitions of the condition number and the pseudospectrum that will
be used throughout this work.

\subsubsection{Condition Number}\label{subsec:condition-number-definition}

Let $B$ be a linear operator with eigenvalue $b_i$. For a simple isolated
eigenvalue, the corresponding right and left eigenvectors, denoted by
$\mathcal{V}_i$ and $\mathcal{U}_i$, are defined by
\begin{equation}
    B^{\dagger} \mathcal{U}_i = b_i^{*}\mathcal{U}_i,
    \qquad
    B \mathcal{V}_i = b_i \mathcal{V}_i,
\end{equation}
where $b_i^{*}$ denotes the complex conjugate of $b_i$. Consider a perturbed
operator of the form
\begin{equation}
    B(\epsilon)=B+\epsilon\,\delta B,
    \qquad
    \|\delta B\|=1.
\end{equation}
Here $\epsilon$ measures the perturbation amplitude. In the numerical results
below, frequencies and operators are expressed in units of the brane scale
$k$; the reported values of $\epsilon$ are therefore dimensionless.

For sufficiently small $\epsilon$, 
one may still solve the perturbed eigenvalue problem
\begin{equation}
    B(\epsilon)\mathcal{V}_i(\epsilon)=b_i(\epsilon)\mathcal{V}_i(\epsilon).
\end{equation}
The sensitivity of $b_i$ to perturbations is then controlled, to leading order,
by the condition number
\begin{equation}
    \label{condition_number_definition}
    |b_i(\epsilon)-b_i|
    \le \epsilon\,\kappa_i+\mathcal{O}(\epsilon^2),
    \qquad
    \kappa_i\equiv \kappa(b_i)
    =\frac{\|\mathcal{U}_i\|\,\|\mathcal{V}_i\|}
    {|\langle \mathcal{U}_i,\mathcal{V}_i\rangle|}.
\end{equation}
For self-adjoint operators, $B=B^\dagger$, the left and right eigenvectors are
proportional, and therefore one has $\kappa_i=1$. In this case the spectrum is
stable in the sense that an operator perturbation of size $\epsilon$ induces, at
most, an eigenvalue shift of the same order. By contrast, for non-self-adjoint
operators the left and right eigenvectors are generally far from being aligned,
and the condition number may become large. As a result, a small
operator perturbation may produce a substantial displacement of the eigenvalue
in the complex plane. 

\subsubsection{Pseudospectrum}\label{subsec:pseudospectrum-definition}

For the finite-dimensional operator obtained after spectral discretization, the
$\epsilon$-pseudospectrum has two equivalent characterizations with respect to
the fixed norm introduced above~\cite{Trefethen:2005}. The perturbative
characterization is

\begin{equation}
    \label{pseudospectrum-definition-1}
    \sigma^\epsilon(B)
    = \left\{ b^\epsilon\in\mathbb{C}:\;\exists\,\delta B
    \ \text{with}\  \|\delta B\|<\epsilon, \ \text{such that}\ 
    b^\epsilon\in\sigma(B+\delta B)
    \right\},
\end{equation}
where $\sigma(\cdot)$ denotes the spectrum of the corresponding operator. This set
definition gives the direct meaning of the pseudospectrum: it collects the
spectra of all perturbations satisfying the norm bound. Eigenvalue trajectories
under a prescribed $\delta B$ are obtained instead from a direct perturbation
calculation.

The second definition is based on the resolvent:
\begin{equation}
    \label{pseudospectrum-definition-2}
    \sigma^\epsilon(B) = \left\{ \lambda\in\mathbb{C}
    :\;  \|R_B(\lambda)\| =  \|(\lambda I-B)^{-1}\| > \epsilon^{-1}
    \right\},
\end{equation}
where $I$ is the identity operator and $R_B(\lambda)$ is the resolvent of $B$;
at spectral points its norm is understood to be infinite. This equivalent
resolvent characterization is computationally useful because it replaces the
generally infeasible task of enumerating all $\delta B$ by a pointwise scan of
the complex $\lambda$-plane. Level sets of the resolvent norm then determine the boundaries
of $\sigma^\epsilon(B)$. Thus Eq.~\eqref{pseudospectrum-definition-1} supplies
the perturbative meaning of the set, whereas
Eq.~\eqref{pseudospectrum-definition-2} supplies the practical computation used
below.

\subsubsection{Energy Inner Product and Norm}\label{subsec:energy-norm}

We now specify the norm used in the numerical analysis. For self-adjoint
problems, one naturally works with the standard $L^2$ inner product in the continuum, or with the
Euclidean inner product after discretization. However, this choice is not well
adapted to the non-self-adjoint QNM problem considered here. A more natural
choice is the energy norm associated with the hyperboloidal evolution
equation~\cite{Jaramillo:2020tuu}.

For two states
\begin{equation}
    \mathcal{V}_1= \binom{\Psi_1}{\Pi_1}, \qquad
    \mathcal{V}_2= \binom{\Psi_2}{\Pi_2}, 
\end{equation}
we define the energy inner product by
\begin{equation}
\langle \mathcal{V}_1,\mathcal{V}_2\rangle_E
= \frac{1}{2} \int_{u_-}^{u_+} \left(
w(u)\,\bar{\Pi}_1\Pi_2 + \xi(u)\,\partial_u\bar{\Psi}_1\,\partial_u\Psi_2
+\hat{V}(u)\,\bar{\Psi}_1\Psi_2 \right)\,du,
\end{equation}
where the overbar denotes complex conjugation. The corresponding norm is
\begin{equation}
    \|\mathcal{V}\|_E =\sqrt{\langle \mathcal{V},\mathcal{V}\rangle_E}.
\end{equation}
For the class of gauges and backgrounds considered here, this inner product is
positive definite and provides the natural measure of the size of perturbations.

The operator norm induced by the energy inner product is then
\begin{equation}
    \|B\|_E = \sqrt{\rho(B^\dagger B)} = 
    \max\left\{\sqrt{b}:\; b\in\sigma(B^\dagger B)\right\},
\end{equation}
where $\rho(\cdot)$ denotes the spectral radius. 
In the corresponding finite-dimensional representation, 
the adjoint with respect to the energy norm is given by
\begin{equation}
    \label{operator_dagger}
    B^\dagger = G_E^{-1}\,B^{*}\,G_E,
\end{equation}
where $B^{*}$ is the Euclidean conjugate transpose and $G_E$ is the Gram
matrix associated with the discrete energy inner product, defined by
\begin{equation}
    \langle \mathcal{V}_1,\mathcal{V}_2\rangle_E = \langle \mathcal{V}_1|G_E|\mathcal{V}_2\rangle.
\end{equation}
Accordingly, the resolvent-based definition of the pseudospectrum can be written
in the equivalent form
\begin{equation}
    \sigma_E^\epsilon(B) 
    = \left\{ \lambda\in\mathbb{C} :\; s_E^{\min}(\lambda I-B)<\epsilon \right\},
\end{equation}
where
\begin{equation}
    s_E^{\min}(M) := \min\left\{\sqrt{m_E}:\; m_E\in\sigma(M^\dagger M)\right\}
\end{equation}
denotes the smallest singular value of $M$ in the energy norm. This quantity
will play a central role in our subsequent pseudospectral analysis.

\subsection{Numerical Approach: Chebyshev Spectral Method}\label{sec:numerical-approach}

To compute the QNM spectrum and the pseudospectrum of the operator $L$,
we employ a Chebyshev spectral method~\cite{Trefethen:2000} on the compactified hyperboloidal domain.
The combination of a finite interval, regular endpoint conditions, and smooth
coefficient functions makes a global spectral discretization effective. For
such coefficients, the Chebyshev method provides rapid convergence and high
numerical accuracy, which is particularly advantageous for resolving
non-self-adjoint spectral structures.

More specifically, the compactified interval $[u_-,u_+]$ is discretized by a
Chebyshev-Lobatto grid, and each component of the state vector
$\mathcal{V}=(\Psi,\Pi)^{\mathsf T}$ is represented by its nodal values on this grid.
Under this discretization, the differential operator $L$ is replaced by a
finite-dimensional matrix $L_N$, obtained by representing derivatives with
Chebyshev differentiation matrices and coefficient functions such as $w(u)$,
$\xi(u)$, $\gamma(u)$, and $\hat V(u)$ by diagonal multiplication matrices.
Accordingly, the continuous spectral problem is reduced to a matrix eigenvalue
problem for $L_N$.
A key advantage of the hyperboloidal formulation is that the outgoing boundary
conditions have already been converted into regularity conditions at the
compactified endpoints. Therefore, at the numerical level one does not need to
impose additional radiative boundary conditions by hand. Instead, it is
sufficient to require regularity of the discrete solution at the endpoints,
which is naturally incorporated into the scheme.

The same discretization also provides the finite-dimensional framework for the
pseudospectral analysis. In the discrete setting, the energy inner product is
represented by a Gram matrix, which induces the corresponding discrete adjoint,
operator norm, and singular values. The quasinormal frequencies are then
obtained from the eigenvalues of $L_N$, while the pseudospectrum is determined
from the resolvent norm, or equivalently from the smallest singular value of
$\lambda I-L_N$ in the discrete energy norm.

In this way, both the QNM problem and the pseudospectral problem are reduced to
standard matrix computations on a finite-dimensional space. As the number of
Chebyshev points $N$ increases, the discrete operator $L_N$ provides an
increasingly accurate approximation to the continuum operator $L$. The
resolution dependence of the computed spectral quantities can therefore be
assessed by repeating the calculation at increasing $N$; representative QNM
convergence tests are presented below.

The Chebyshev differentiation matrices follow the standard collocation
framework~\cite{Trefethen:2000}. The explicit construction of the discrete
hyperboloidal operator and the associated Gram matrix is presented in the
appendix of Ref.~\cite{Jia:2026vnx}.

\subsection{Validation of the Characteristic Spectrum}\label{sec:validation-spectrum}

\begin{figure*}[htb]
    \begin{center}
    \includegraphics[width=8.0cm]{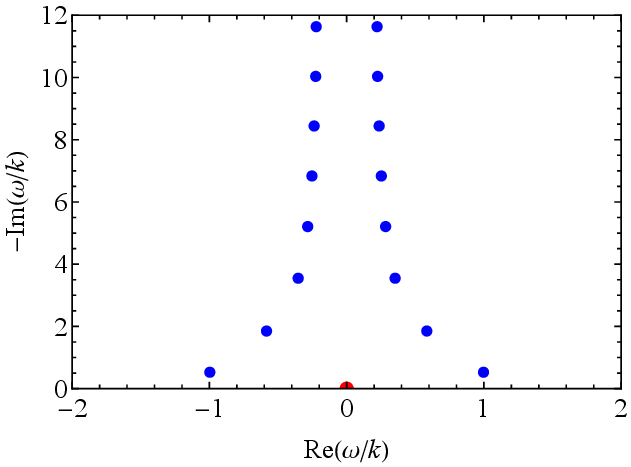}
    \end{center}
    \caption{QNM spectrum (blue dots) and the zero mode (red dot)
    for gravitational perturbations of the representative thick brane.}\label{figure-2}
\end{figure*}

As a first validation of the hyperboloidal framework 
and spectral discretization developed above, we recompute the QNM spectrum of
the thick brane system.
The resulting frequencies are in close agreement with those reported in
Ref.~\cite{Tan:2022vfe}, supporting the numerical implementation.

\begin{figure*}[htb]
    \begin{center}
    \subfigure{\label{spectralmethod-error-n1}
    \includegraphics[width=7.6cm]{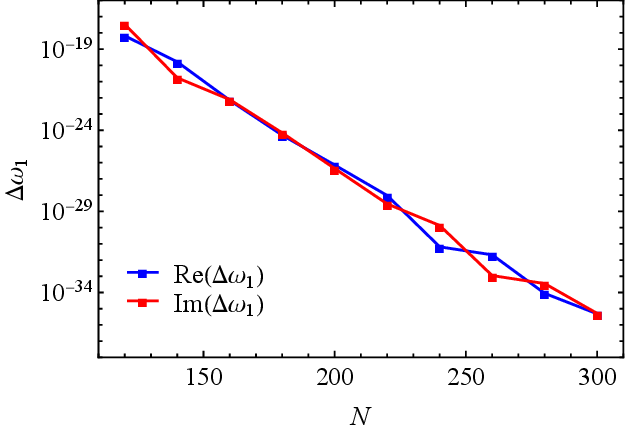}}
    \subfigure{\label{spectralmethod-error-n3}
    \includegraphics[width=7.6cm]{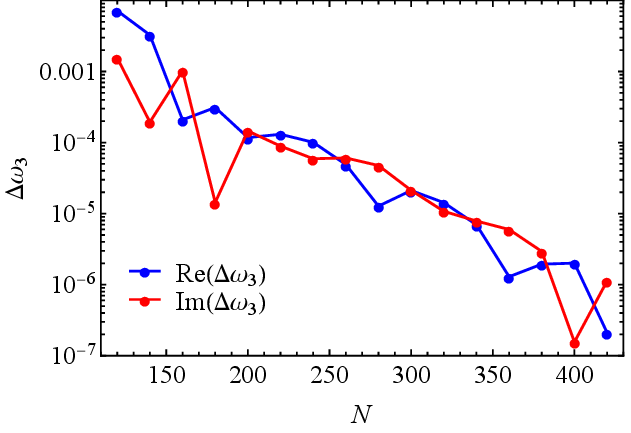}}
    \end{center}
    \caption{Componentwise resolution-convergence errors for the
    fundamental QNM (left panel) and third QNM (right panel), with $\Delta N=20$.
    The blue and red curves show $\Delta\omega_n^{(\mathrm{Re})}$ and
    $\Delta\omega_n^{(\mathrm{Im})}$, respectively.}\label{figure-error}
\end{figure*}

Figure~\ref{figure-2} displays the discrete QNM spectrum (blue dots) obtained for $N=500$,
after excluding the resolution-dependent sequence of points
along the imaginary axis that represents the discretized continuum or branch-cut
structure.
To further assess the numerical robustness of the method,
Fig.~\ref{figure-error} presents a resolution-convergence analysis.
We define the componentwise error estimates as
\begin{equation}
    \begin{aligned}
    \Delta\omega_n^{(\mathrm{Re})}(N)
    &=\left|\mathrm{Re}\,\omega_n(N+\Delta N)
      -\mathrm{Re}\,\omega_n(N)\right|,\\
    \Delta\omega_n^{(\mathrm{Im})}(N)
    &=\left|\mathrm{Im}\,\omega_n(N+\Delta N)
      -\mathrm{Im}\,\omega_n(N)\right|.
    \end{aligned}
\end{equation}
Here $\omega_n(N)$ denotes the $n$th QNM frequency computed at
spectral resolution $N$.

Both componentwise errors decrease overall as $N$ increases. The fundamental
QNM exhibits rapid, nearly monotonic convergence over the displayed range,
whereas the third QNM converges more slowly and nonmonotonically. Nevertheless,
the errors of the third QNM still decrease by several orders of magnitude.
These trends support the numerical reliability of
the frequencies used below.

Note that the mode located at the origin in Fig.~\ref{figure-2}, 
marked by the red point, is not a quasinormal mode. 
Rather, it corresponds to the localized zero mode for the gravitational perturbations
of the thick brane, as established in the analytical discussion above.

\subsection{Condition Numbers of the Characteristic Modes}\label{sec:condition-numbers}

\begin{figure*}[htb]
    \begin{center}
    \includegraphics[width=8.0cm]{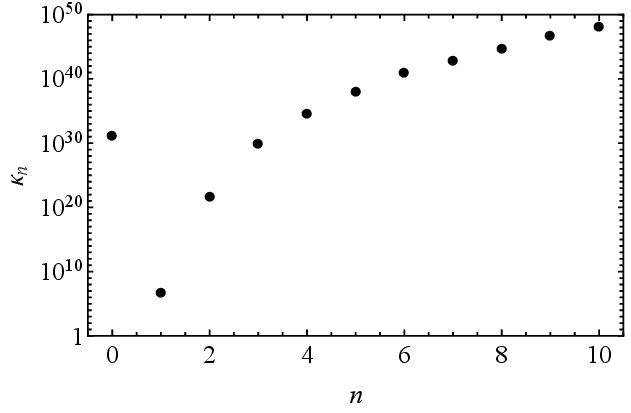}
    \end{center}
    \caption{Condition numbers $\kappa_n$ of the characteristic spectrum.}\label{figure-3}
\end{figure*}

We now examine the spectral sensitivity of the characteristic modes using the
two complementary diagnostics introduced in Sec.~\ref{sec:definitions}. We
begin with the condition number, which quantifies the local first-order response
of an individual eigenvalue. Using the energy-norm adjoint in
Eq.~\eqref{operator_dagger}, we compute the left eigenvectors
$\mathcal{U}_i$ and combine them with the corresponding right eigenvectors
$\mathcal{V}_i$ through Eq.~\eqref{condition_number_definition}. The resulting
condition numbers are shown in Fig.~\ref{figure-3}.

As shown in Fig.~\ref{figure-3}, the condition numbers $\kappa_n$ of the
QNM overtones increase rapidly with the overtone number $n$. This indicates
that higher overtones are more sensitive to perturbations than the fundamental
QNM $\omega_1$, and that this sensitivity becomes progressively stronger along
the QNM sequence.

A noteworthy feature is that the zero mode condition number $\kappa_0$ is much
larger than that of the fundamental QNM. The zero mode therefore has a
stronger local first-order response to generic perturbations of the
hyperboloidal operator in the chosen energy norm. This operator-level statement
is distinct from the bound-state result derived from the factorized
Schr\"odinger operator. Self-consistent background deformations obey the coupled
braneworld equations and form a restricted perturbation class; their effect on
the zero mode must therefore be assessed within that class.

\subsection{Pseudospectral Analysis}\label{sec:pseudospectral-analysis}

Condition numbers describe the local first-order sensitivity of individual
eigenvalues. Their connection with the local pseudospectral geometry follows
from the pole of the resolvent. For a simple isolated eigenvalue $\omega_n$ and
for $\omega$ sufficiently close to that eigenvalue,
\begin{equation}
    \label{local-resolvent-condition}
    \left\|R_{L_N}(\omega)\right\|_E
    \simeq \frac{\kappa_n}{|\omega-\omega_n|},
    \qquad
    s_E^{\min}(\omega I-L_N)
    \simeq \frac{|\omega-\omega_n|}{\kappa_n}.
\end{equation}
Consequently, the local boundary of the $\epsilon$-pseudospectrum satisfies
$|\omega-\omega_n|\simeq\epsilon\kappa_n$ while the eigenvalue remains
isolated. The condition number therefore predicts the leading local scale,
whereas the pseudospectrum captures the subsequent deformation and merger of
these neighborhoods at finite $\epsilon$.

We compute the pseudospectrum from the resolvent
characterization~\eqref{pseudospectrum-definition-2}. Specifically, we evaluate
$s_E^{\min}(\omega I-L_N)$ pointwise over the complex-frequency plane. The
level set $s_E^{\min}(\omega I-L_N)=\epsilon$ is the boundary of
$\sigma_E^\epsilon(L_N)$, and the region satisfying
$s_E^{\min}<\epsilon$ belongs to the pseudospectrum. Equivalently, the
pointwise value
$\epsilon_\star(\omega)\equiv s_E^{\min}(\omega I-L_N)$ is the threshold
above which $\omega$ belongs to the $\epsilon$-pseudospectrum.

\begin{figure*}[htb]
    \begin{center}
    \subfigure{\label{pseudospectrum_region_eps_1e-8}
    \includegraphics[width=7.6cm]{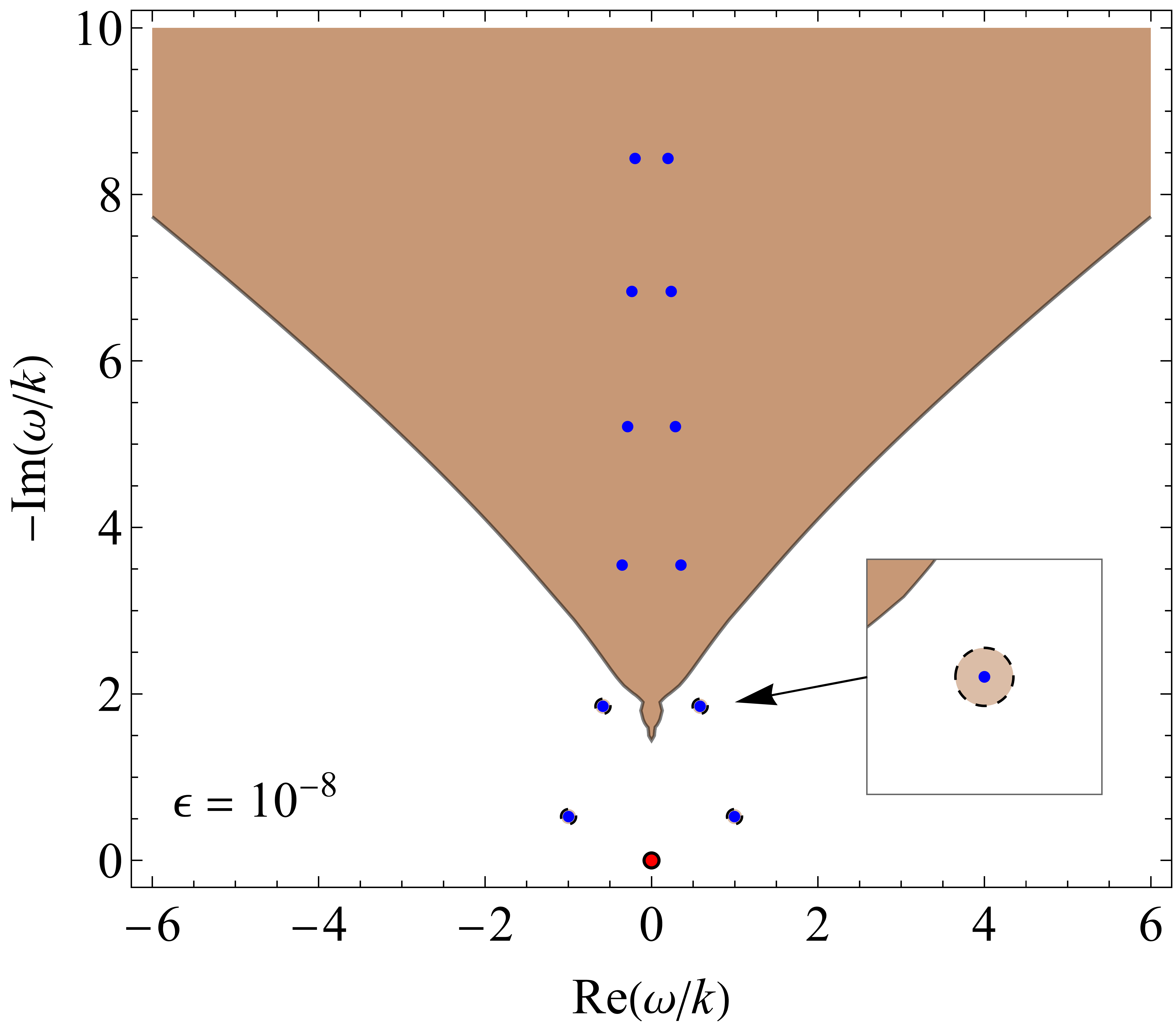}}
    \subfigure{\label{pseudospectrum_region_eps_1e-4}
    \includegraphics[width=7.6cm]{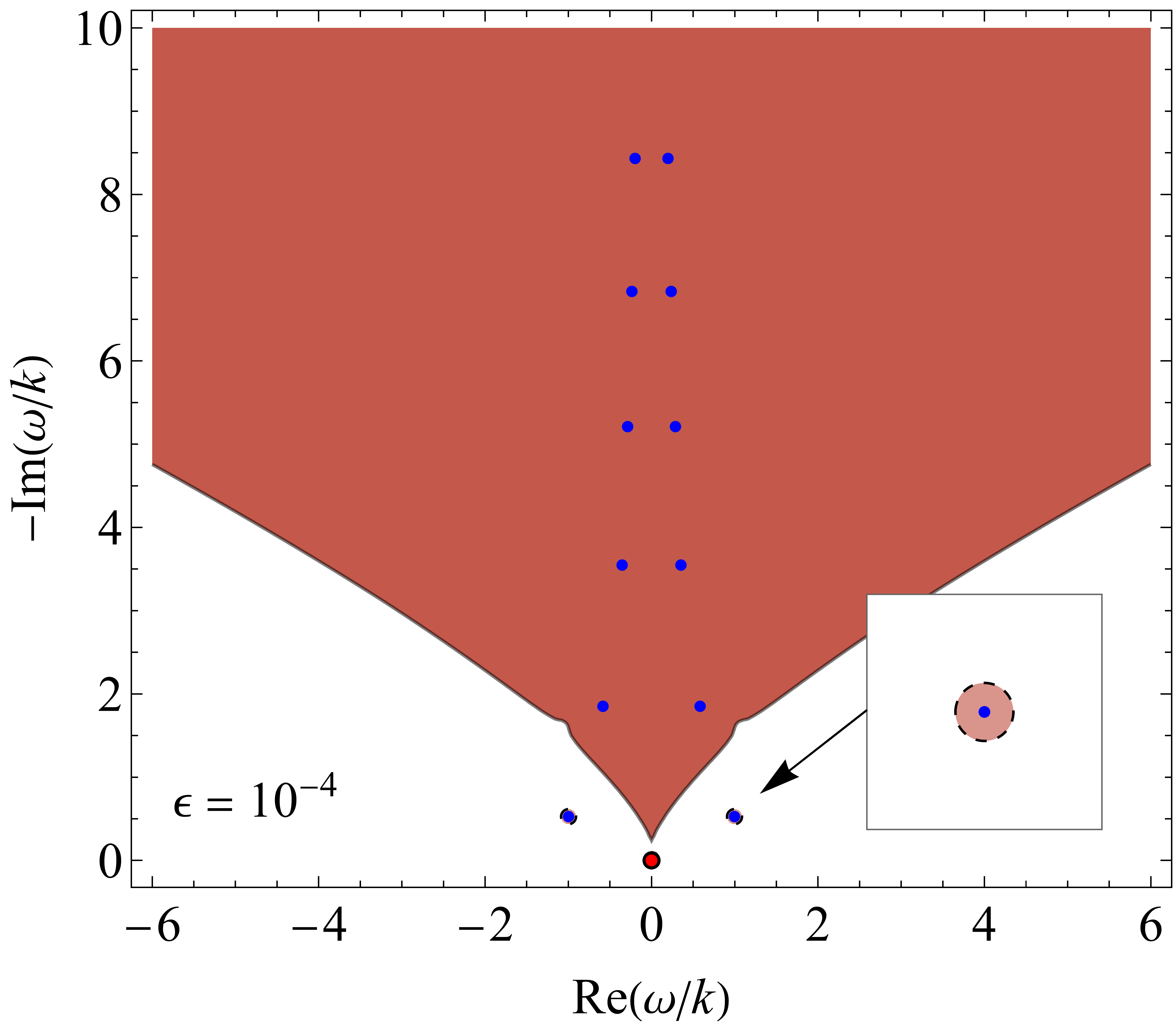}}
    \end{center}
    \caption{%
    $\epsilon$-pseudospectra of the thick brane. The shaded regions denote
    $\sigma_E^\epsilon(L_N)$ computed with respect to the energy norm, while the
    blue dots mark the QNM frequencies and the red dot denotes the localized zero
    mode. The left and right panels correspond to $\epsilon=10^{-8}$ and
    $\epsilon=10^{-4}$, respectively. The insets show a magnified view of the
    pseudospectral boundary around a representative QNM.}\label{figure-4}
\end{figure*}

Figure~\ref{figure-4} shows $\sigma_E^\epsilon(L_N)$ at the two representative
levels $\epsilon=10^{-8}$ and $10^{-4}$. Since
$\sigma(L_N)\subset\sigma_E^\epsilon(L_N)$ for every $\epsilon>0$, the relevant
information is not whether an eigenvalue appears in the set, but how its local
component enlarges and merges with neighboring components. At
$\epsilon=10^{-8}$, the local components surrounding the higher overtones shown
in the figure have already merged into a broad connected domain, whereas
several lower overtones remain isolated. When $\epsilon$ is increased to
$10^{-4}$, this connected domain extends toward the lower-overtone region, in
accordance with the nesting property
$\sigma_E^{\epsilon_1}(L_N)\subseteq\sigma_E^{\epsilon_2}(L_N)$ for
$\epsilon_1<\epsilon_2$. The earlier merger of the higher-overtone components
is consistent with the rapid growth of their condition numbers.

The fundamental QNM and the localized zero mode remain in separate local
components at both displayed levels. This separation from the high-overtone
domain is a statement about global connectivity; the local sensitivity of each
isolated mode is quantified by its own condition number and local contours.

\begin{figure*}[htb]
    \begin{center}
    \subfigure{\label{pseudospectrum_region}
    \includegraphics[width=8.8cm]{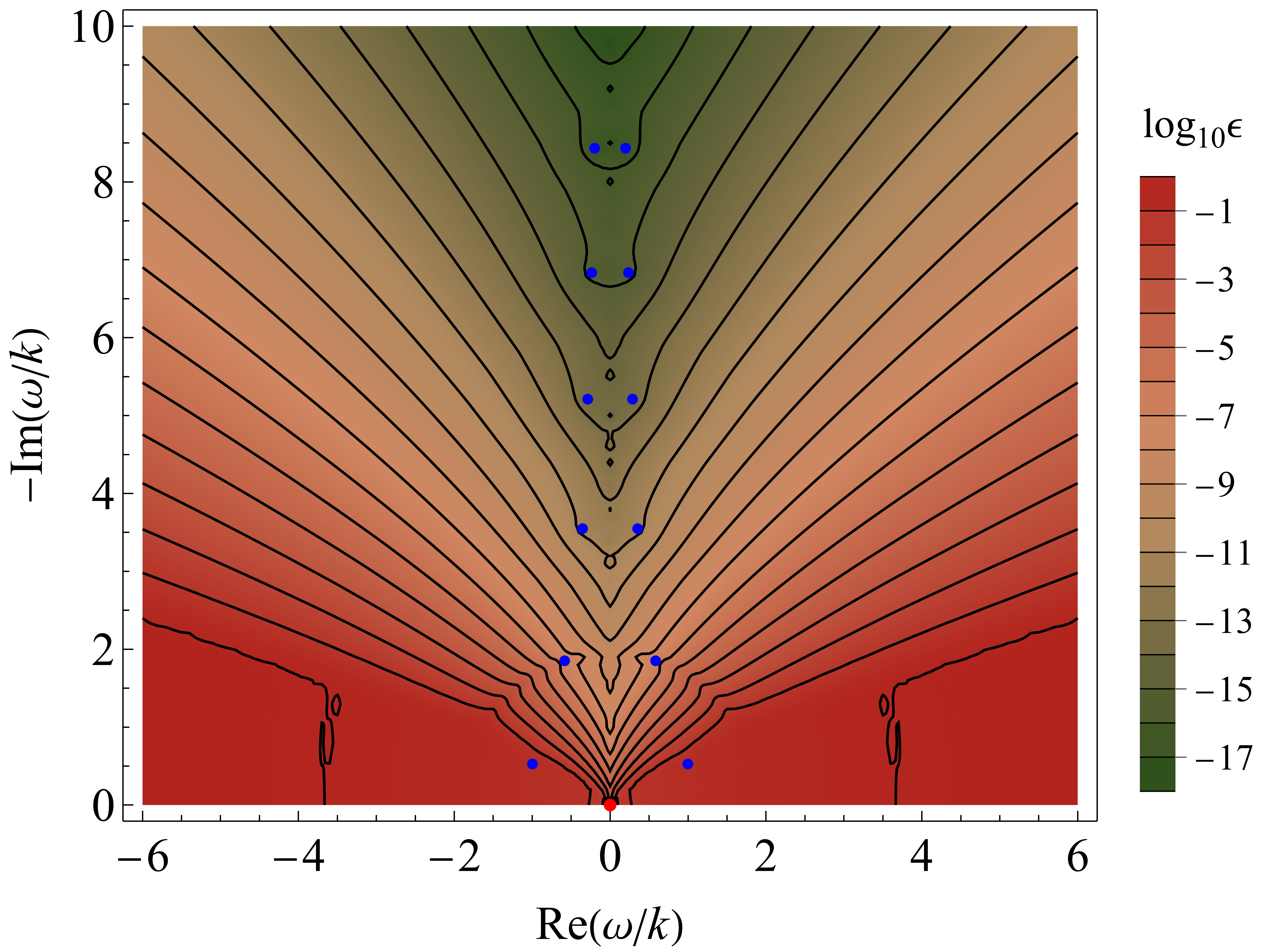}}
    \subfigure{\label{pseudospectrum_region_local}
    \includegraphics[width=7.4cm]{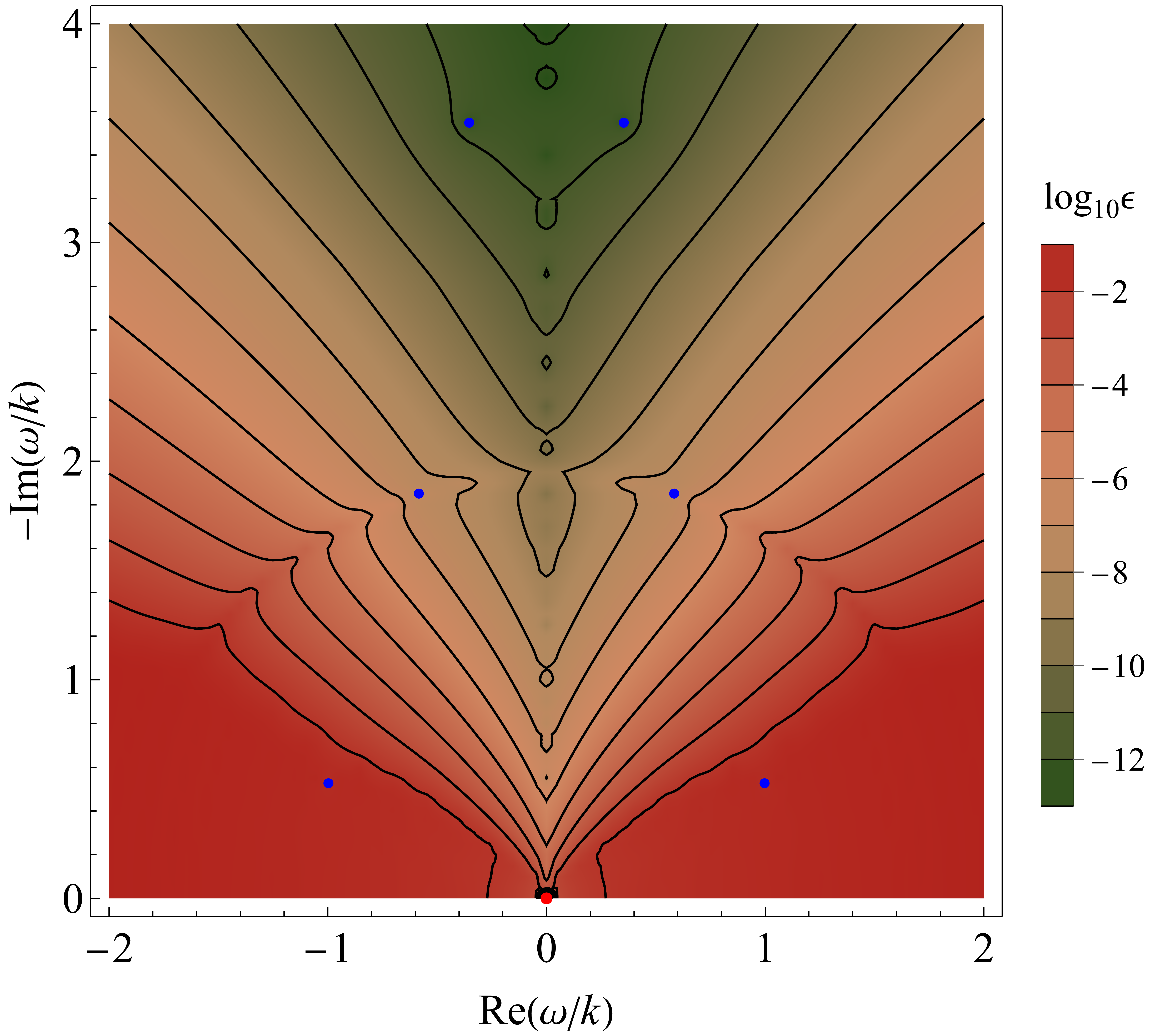}}
    \end{center}
    \caption{%
    Pseudospectral contour plots of the thick brane. 
    The left panel shows the global pseudospectral structure in the complex
    frequency plane, while the right panel gives a magnified view of the
    low-overtone region near the zero mode. 
    The color scale displays $\log_{10}s_E^{\min}(\omega I-L_N)$, where
    $s_E^{\min}$ is the smallest singular value in the energy norm; its label
    $\log_{10}\epsilon$ refers to the contour value
    $\epsilon=s_E^{\min}(\omega I-L_N)$. More negative values correspond to a
    smaller perturbation threshold and hence to stronger spectral sensitivity.
    The black curves are constant-$\epsilon$ contours and give the
    boundaries of the corresponding $\epsilon$-pseudospectra
    $\sigma_E^\epsilon(L_N)$. 
    The blue dots denote the QNM frequencies, and the red dot marks the localized
    zero mode.}\label{figure-5}
\end{figure*}

To obtain a broader view, we further scan the region
$\mathrm{Re}(\omega/k)\in[-6,6]$ and $-\mathrm{Im}(\omega/k)\in[0,10]$ 
in the complex frequency plane. 
The resulting contour plot is shown in Fig.~\ref{figure-5}.

\begin{figure*}[!t]
    \centering
    \subfigure[~Zero mode]{\label{pseudospectrum_region_zeromode}
    \includegraphics[width=8.4cm]{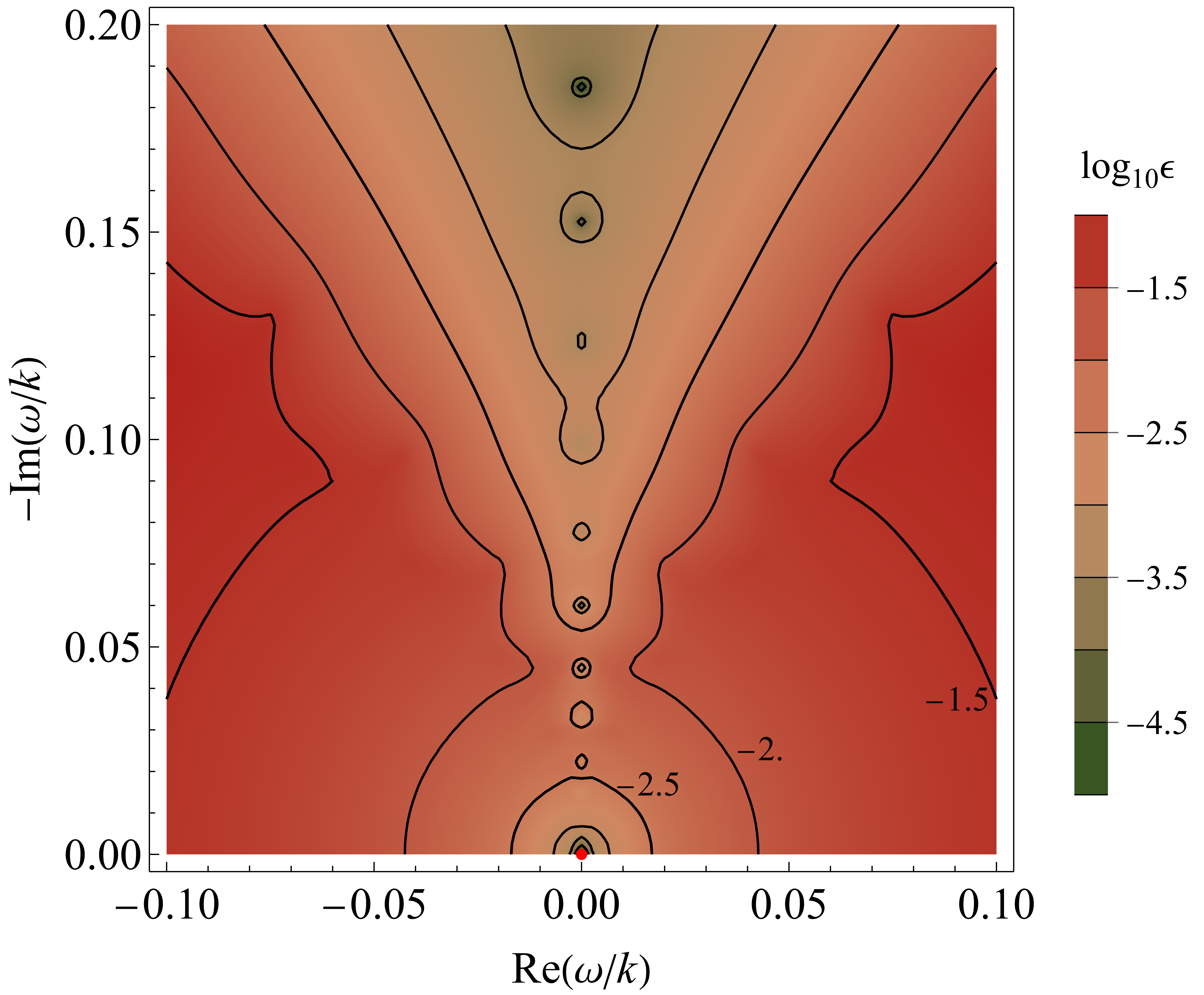}}
    \subfigure[~Fundamental QNM]{\label{pseudospectrum_region_Omega1}
    \includegraphics[width=8.4cm]{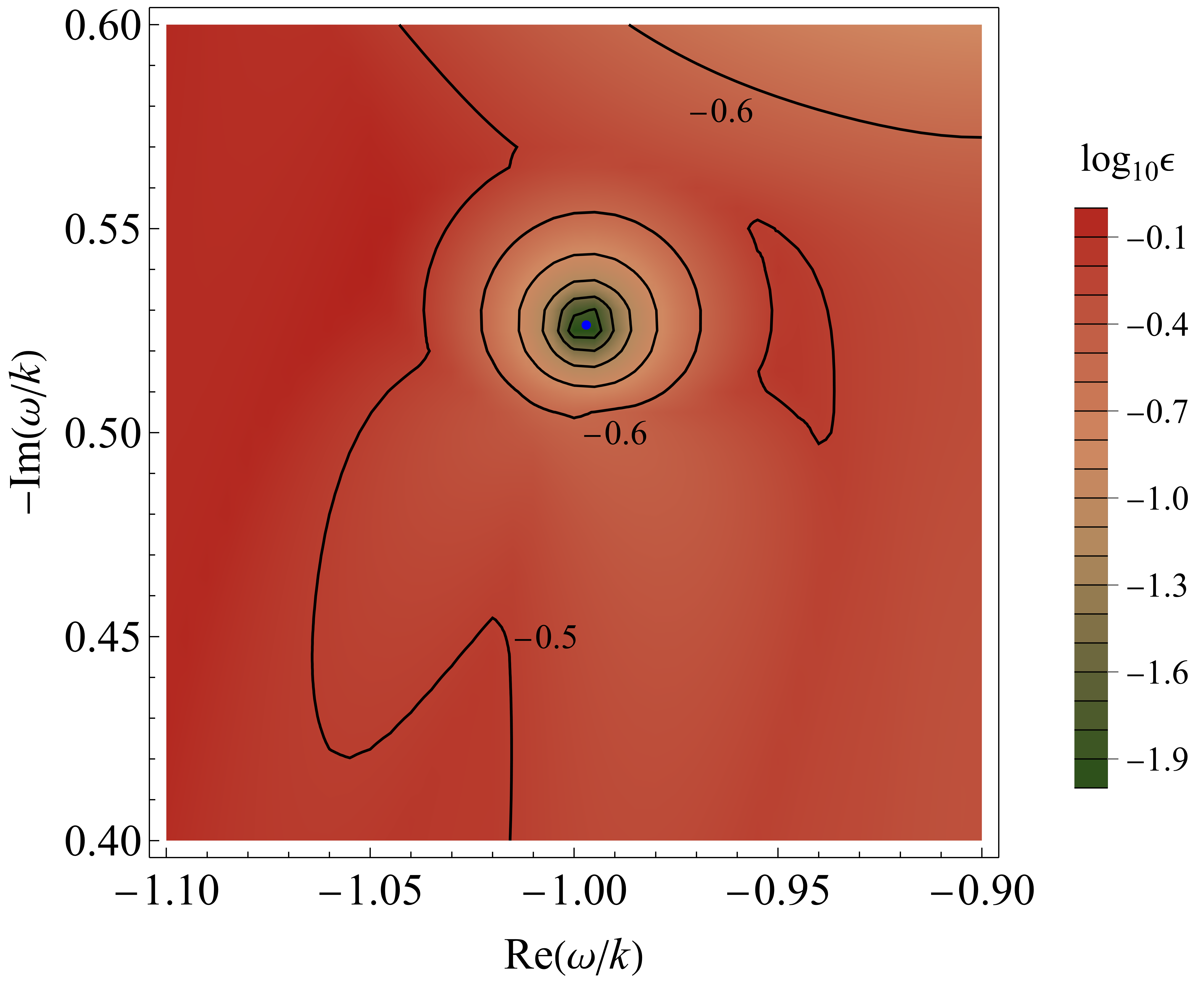}}\\
    \subfigure[~Second QNM]{\label{pseudospectrum_region_Omega2}
    \includegraphics[width=8.4cm]{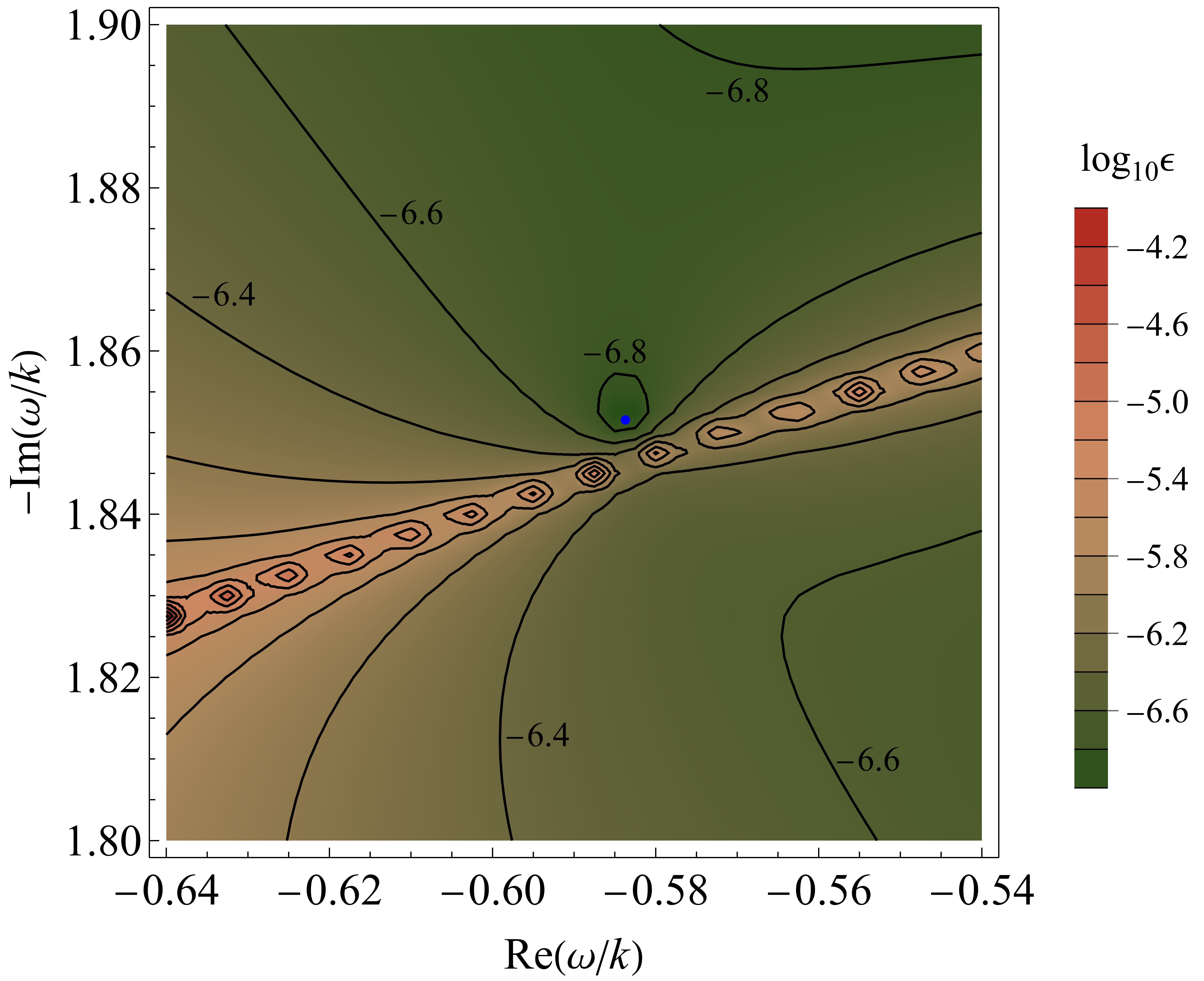}}
    \subfigure[~Third QNM]{\label{pseudospectrum_region_Omega3}
    \includegraphics[width=8.4cm]{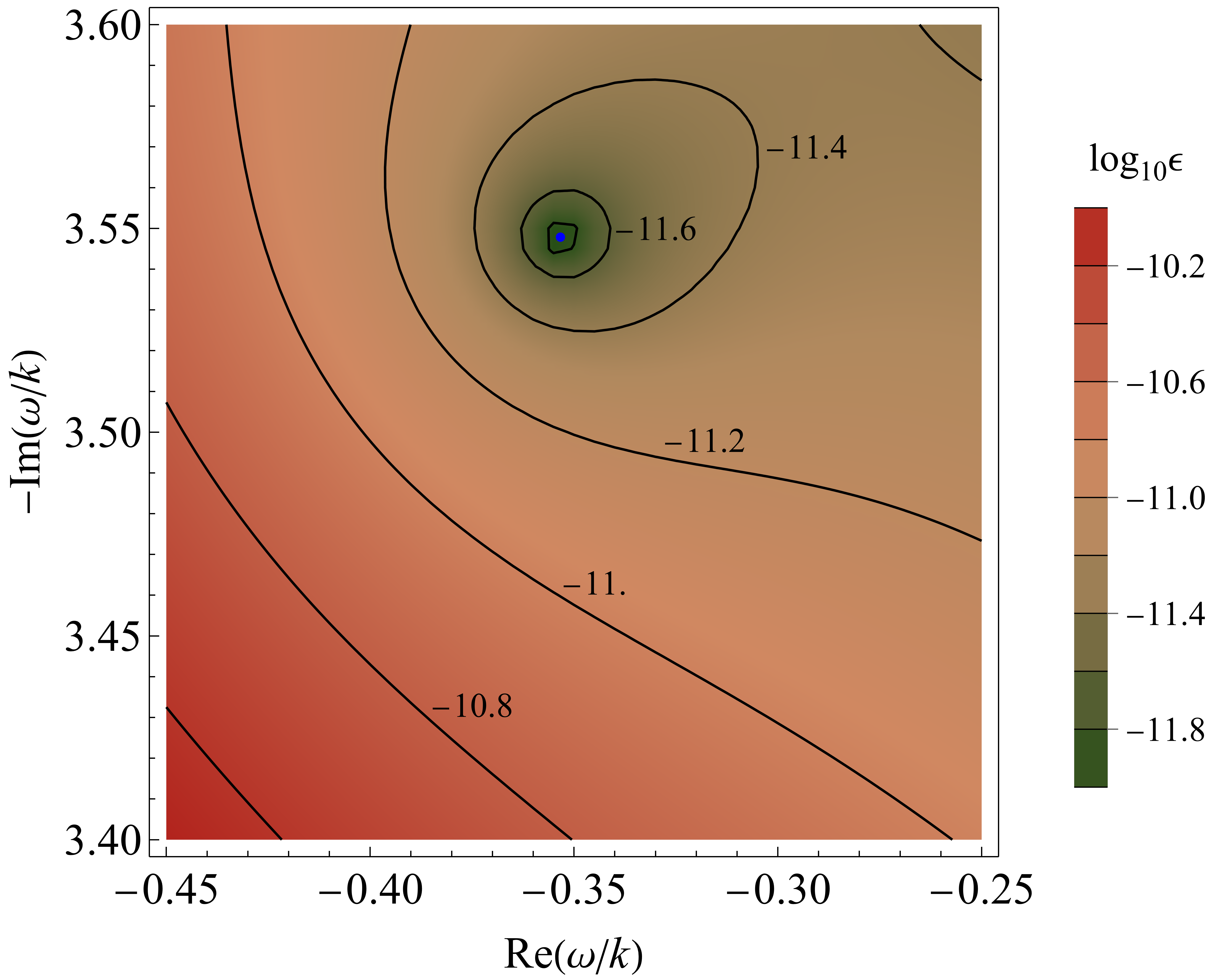}}
    \caption{
    Local views of the smallest-singular-value field shown in
    Fig.~\ref{figure-5}. Panels (a)--(d) are centered near the localized zero
    mode $\omega_0=0$, the fundamental QNM $\omega_1$, the second QNM $\omega_2$,
    and the third QNM $\omega_3$, respectively. The color scale displays
    $\log_{10}s_E^{\min}(\omega I-L_N)$; its label $\log_{10}\epsilon$ denotes
    the corresponding contour value, with more negative values indicating a
    smaller perturbation threshold. The black curves are
    constant-$\epsilon$ boundaries of $\sigma_E^\epsilon(L_N)$. Each panel uses
    its own plotting window and color range. Contour-shape comparisons are
    therefore qualitative, and each $\epsilon$ value must be read from the scale
    of its own panel. Blue dots mark QNM frequencies, and the red dot marks the
    localized zero mode.}\label{figure-6}
\end{figure*}

Figure~\ref{figure-5} displays the pointwise field
$\log_{10}\epsilon_\star(\omega)$. More negative values identify frequencies
that can enter the pseudospectrum under smaller operator perturbations. The
global panel reveals a low-threshold valley following the high-overtone QNM
sequence. As $-\mathrm{Im}(\omega/k)$ increases, the level curves become more
elongated and the neighborhoods of neighboring overtones overlap, producing the
connected component seen in Fig.~\ref{figure-4}. This anisotropy shows that the
spectral response is strongly direction dependent, as expected for a non-normal
evolution operator. The right panel magnifies the transition between the
isolated low-lying modes and this extended high-overtone structure and locates
the regions examined in Fig.~\ref{figure-6}.

Figure~\ref{figure-6} resolves the local contour geometries hidden by the scale
of the global panel. All four panels sample the same smallest-singular-value
field, but use different windows and color ranges; their contour areas should
therefore not be compared directly. At the finest displayed levels, a closed
component around an isolated eigenvalue has the local interpretation given by
Eq.~\eqref{local-resolvent-condition}. As $\epsilon$ increases, its deformation
and eventual merger with the surrounding field show how an isolated local
neighborhood gives way to collective pseudospectral structure. After such a
merger, the connected set describes possible spectral locations over the full
norm-bounded perturbation class rather than a unique continuation of one mode.

Accordingly, three features of the local plots carry different information.
First, the distance from $\omega_n$ to an innermost contour at fixed
$\epsilon$ measures the local amplification scale and can be compared with
$\epsilon\kappa_n$ in the isolated-pole regime. Second, a nearly circular,
nested family is the geometry expected when the pole term dominates, whereas
pronounced elongation shows that the regular part of the resolvent and/or nearby
spectral components already influence the response, making it direction
dependent. Third, the value of $\epsilon$ at which a closed component joins the
surrounding valley identifies the crossover beyond which a mode-by-mode local
description is no longer sufficient. These three diagnostics---local size,
anisotropy, and merger scale---provide the basis for the panel-by-panel reading
below.

Since $\epsilon_\star(\omega_n)=0$ at every exact eigenvalue, the point color
itself cannot rank stability. Instead,
Eq.~\eqref{local-resolvent-condition} gives the local radial slope
$\partial s_E^{\min}/\partial|\omega-\omega_n|\simeq1/\kappa_n$: a larger
$\kappa_n$ produces a shallower low-threshold basin and a wider fixed-$\epsilon$
contour, while direction-dependent slopes mark departure from the isolated-pole
geometry. We therefore compare numerical contour labels and merger levels, not
raw colors across independently scaled panels.

\begin{itemize}
    \item 
    In the zero mode panel, the innermost component encloses $\omega=0$, while
    the outer contours are elongated along the imaginary direction. Together
    with $\kappa_0>\kappa_1$ in Fig.~\ref{figure-3}, this shows that the zero mode
    has a stronger local first-order response than the fundamental QNM in the
    chosen energy norm. The additional small closed components aligned with the
    imaginary axis are likely associated with the finite-resolution
    representation of the continuum or branch-cut structure. A definitive
    identification of these components would require tracking their behavior as
    the spectral resolution is increased.
    
    \item 
    The fundamental QNM panel contains a regular family of nested contours
    around $\omega_1$ and remains well separated from the extended structure at
    the displayed levels. This compact local geometry is consistent with its
    comparatively small condition number. Among the damped modes shown, the
    fundamental QNM is therefore the most robust against generic operator
    perturbations in the chosen norm.
    
    \item 
    Near the second QNM, the level curves are strongly anisotropic and coexist
    with a narrow chain of small components directed toward the high-overtone
    valley. The finest contour around $\omega_2$ still defines an isolated local
    neighborhood, whereas the coarser contours approach the neighboring chain
    and the extended background. This is the first low-overtone panel in which
    the transition from an isolated mode to the collective high-overtone
    structure becomes pronounced.
    
    \item 
    Around the third QNM, the finest displayed contours still form a local
    component enclosing $\omega_3$, whereas coarser contours are embedded in an
    anisotropic background that is continuous with the high-overtone valley in
    Fig.~\ref{figure-5}. The loss of a clearly separated neighborhood therefore
    occurs at a finer perturbation scale than for the fundamental QNM, in
    qualitative agreement with the larger condition number of the third mode.
    
\end{itemize}

Taken together, Figs.~\ref{figure-4}--\ref{figure-6} establish a consistent
hierarchy among the damped modes shown. The fundamental QNM has the most compact
local contours and the smallest condition number, while increasing overtone
number is accompanied by larger $\kappa_n$, lower perturbation thresholds,
stronger anisotropy, and earlier merger into the connected high-overtone
domain. As $\epsilon$ increases, this connected domain expands from the
high-overtone sector toward the lower overtones. The zero mode is exceptional:
its local sensitivity exceeds that of the fundamental QNM, although its
physical persistence is additionally controlled by the factorized tensor
operator and the constraints on self-consistent braneworld deformations. Thus
the present pseudospectrum gives the operator-level sensitivity hierarchy in
the energy norm; a deformation-specific mode trajectory requires fixing a
particular admissible perturbation.

\section{Conclusion}\label{sec:conclusion}

In this paper, we applied pseudospectral analysis to gravitational perturbations
of a representative thick brane. For the transverse-traceless tensor sector, we
combined a hyperboloidal compactification with a first-order formulation of the
master wave equation. The outgoing QNM conditions then become regularity
conditions at the compactified endpoints, and the problem is governed by a
non-self-adjoint evolution operator $L$ on a finite interval. Using the energy
norm of the hyperboloidal wave equation, we constructed the corresponding
adjoint, condition numbers, smallest singular values, and
$\epsilon$-pseudospectra, and implemented the calculation with a Chebyshev
spectral discretization.

The localized graviton zero mode and the QNM spectrum were recovered, and the
numerical convergence of representative QNMs was verified as the spectral
resolution was increased.
The condition numbers quantify the local first-order response of individual
spectral points to generic perturbations of the hyperboloidal operator. They grow
rapidly with the QNM overtone number, showing that the higher overtones are
locally more sensitive than the fundamental QNM in the chosen energy norm. The
zero mode also has a larger condition number than the fundamental QNM.

The pseudospectrum supplies the corresponding finite-perturbation and global
picture. Near a simple isolated eigenvalue, the local contour scale satisfies
$|\omega-\omega_n|\simeq\epsilon\kappa_n$, directly connecting the contour
geometry with the condition-number hierarchy. At larger $\epsilon$, contour
deformation and merger reveal collective non-normal response beyond this local
approximation. The global field contains a low-threshold valley along the
high-overtone sequence: the high-overtone components merge first, and the
connected domain expands toward the lower overtones as $\epsilon$ increases.
The fundamental QNM retains the most compact local structure among the damped
modes shown. The second QNM marks the onset of a pronounced chain-like,
anisotropic structure, while the third QNM is already embedded in the extended
high-overtone background. Each $\epsilon$-pseudospectrum contains the possible
spectral locations over the full norm-bounded perturbation class; a prescribed
deformation instead selects particular mode trajectories.

The zero mode provides a distinct structural comparison. Its larger condition
number and elongated local contours show a stronger generic operator-level
response than that of the fundamental QNM. Its persistence under physical thick
brane deformations is additionally governed by the factorized tensor operator
and by the coupled equations constraining self-consistent backgrounds. The
unrestricted pseudospectrum therefore provides a reference for
more specific tests within a structure-preserving perturbation class.

These results establish pseudospectral methods as a useful diagnostic of the
spectral sensitivity of braneworld characteristic modes. They complement direct
perturbative calculations by revealing the worst-case response over a
norm-bounded operator class. A natural next step is to construct explicit
structure-preserving perturbations from self-consistent thick brane backgrounds
and determine which parts of the unrestricted pseudospectrum are physically
realizable. It will also be useful to investigate whether the high-overtone
sensitivity has a time-domain signature and to extend the framework to other
braneworld geometries and perturbation sectors.

\section*{Acknowledgments}

We would like to thank Wen-Yi Zhou for very useful discussions. 
This work was supported by 
the National Natural Science Foundation of China (Grant Nos. 12475056, 12205129, and 12247101),
the Fundamental Research Funds for the Central Universities (Grant Nos. lzujbky-2025-it05 and lzujbky-2025-jdzx07), 
the Natural Science Foundation of Gansu Province (Grant Nos. 22JR5RA389 and 25JRRA799), 
Gansu Province's Top Leading Talent Support Plan, 
and the ``111 Center'' under Grant No. B20063. 
Wen-Di Guo was supported by ``Talent Scientific Fund of Lanzhou University''.
\par

\end{document}